\documentclass[graybox]{svmult}

\usepackage{mathptmx}       % selects Times Roman as basic font
\usepackage{helvet}         % selects Helvetica as sans-serif font
\usepackage{courier}        % selects Courier as typewriter font
\usepackage{type1cm}        % activate if the above 3 fonts are
\usepackage{makeidx}         % allows index generation
\usepackage{graphicx}        % standard LaTeX graphics tool
\usepackage{multicol}        % used for the two-column index
\usepackage[bottom]{footmisc}% places footnotes at page bottom
\usepackage{amsmath}
\usepackage{bm}
\usepackage{tikz}

\makeindex             % used for the subject index
\newcommand{\condip}{\mathrel{\text{\scalebox{1.07}{$\perp\mkern-10mu\perp$}}}}
\newcommand{\given}{\,|\,}

\begin{document}

\title*{Customised Structural Elicitation}
% Use \titlerunning{Short Title} for an abbreviated version of
% your contribution title if the original one is too long
\author{Rachel L. Wilkerson and Jim Q. Smith}
% Use \authorrunning{Short Title} for an abbreviated version of
% your contribution title if the original one is too long
\institute{Rachel L. Wilkerson \at University of Warwick, \email{R.L.Wilkerson@warwick.ac.uk}
\and Jim Q. Smith \at University of Warwick, Alan Turing Institute \email{J.Q.Smith@warwick.ac.uk}}
\maketitle

\abstract*{Established methods for structural elicitation typically rely on code modelling standard graphical models classes, most often Bayesian networks. However, more appropriate models may arise from asking the expert questions in common language about what might relate to what and exploring the logical implications of the statements. Only after identifying the best matching structure should this be embellished into a fully quantified probability model. Examples of the efficacy and potential of this more flexible approach are shown below for four classes of graphical models: Bayesian networks, Chain Event Graphs, Multi-regression Dynamic Models, and Flow Graphs. We argue that to be fully effective any structural elicitation phase must first be customised to an application and if necessary new types of structure with their own bespoke semantics elicited.}

\section{Background}
\label{sec:background}
Expert elicitation is a powerful tool when modelling complex problems especially in the common scenario where current probabilities are unknown and data is unavailable for certain regions of the probability space. Such methods are now widely developed, well understood, and have been used to model systems in a variety of domains including climate change, food insecurity, and nuclear risk assessment \cite{Barons2018,Rougier2014,Hanea2007}.
However, eliciting expert probabilities faithfully has proved to be a sensitive task, particularly in multivariate settings. We argue that first eliciting structure is critical to the accuracy of the model, particularly as conducting a probability elicitation is time and resource-intensive. 

An appropriate model structure fulfils two criteria. Firstly, it should be compatible with how experts naturally describe a process. Ideally, modellers should agree upon a structure using natural language. Secondly, any structure should ideally have the potential to eventually be embellished through probabilistic elicitation into  a full probability model.  It is often essential to determine that the structure of a problem as desired by a client is actually consistent with the class of structural models considered. The logic and dynamics of Bayesian networks often do not match with an experts' description of a problem. When this happens, a customising approach as we illustrate below generates flexible models that are a more accurate representation of the process described by the domain expert. We show that these alternative graph models often admit a supporting formal framework and subsequent probabilistic model similar to a BN while more faithfully representing the beliefs of the experts.

While there are several protocols for eliciting probabilities such as the Cooke method, SHELF and IDEA protocols, the process of determining the appropriate underlying structure has not received the same attention.  Protocols for eliciting structural relationships between variables in the continuous range have been developed \cite{tim2001probabilistic} and basic guidelines for eliciting a discrete Bayesian network structure are available and well documented \cite{Korb2009,JQSdecisionAnalysis}. These methods are widely applicable, but are rarely customised to structural elicitation of frameworks other than those provided by the BN. However, it is possible to develop customising protocols to elicit structure. We illustrate this through the case studies in this paper.

Towards this end, this paper explores examples of real case studies that are better-suited  to eliciting bespoke structure. We illustrate experts' natural language description of a problem can determine the structure of a model. Programs to alleviate food insecurity in the United States serves as a running example. Even within this domain we are able to show that  different problem dynamics are naturally more suited to particular structures, and eliciting these custom structures creates more compelling models. We show that these bespoke structures can subsequently be embellished into customised probabilistic graphical models that support a full probabilistic description. 

\section{Eliciting model structure}
\label{sec:structure}
Structured expert elicitation begins with a natural language description of the problem from domain experts. An expert describes the components of a system and how they are related, and a structure often emerges organically. This process may be aided by the use of informal graphs, a widespread practice. However, the methods used by the facilitators for systematically translating these diagrams into their logical consequences and finally embellishing these into a full probability model is often not supported. Nevertheless, there are certain well developed classes of graphical models that do support this translation. The most popular and best supported by available software is the BN. However, other graphical frameworks have emerged, each with its own representative advantages. These include event trees, chain event graphs, and dynamic analogues of theses \cite{CEGbook, Barclay2015}. We describe some of the competing frameworks and suggest how one can be selected over another.

\subsection{Choosing an appropriate structure}
\label{sec:pickstruct}
Choosing between candidate structures may not be straightforward. Some domain problems may be compatible with existing structures, while others might require creating new classes of graphical models. The task of developing a bespoke graphical framework that supports a translation into a choice of probability models is usually a labour-intensive one requiring some mathematical skills. 
While some domain problems will require the modeller to undertake developing a customised model class, there are also several such frameworks already built, forming a tool-kit of different frameworks. We give guidelines below to help the modeller decide which of these methods most closely matches the problem explanation evolving from the natural language of the experts.

As a running example, we consider the drivers of food insecurity. The illustrations we use throughout the paper are based on meetings with actual domain experts. We have simplified these case studies so that we can illustrate the elicitation process as clearly as possible. A meeting of advocates discusses the effect of food insecurity on long-term health outcomes. One advocate voices that food insecurity stems from insufficient resources to purchase food. The experts collectively attest that the two main sources of food are personal funds like disposable income or government benefit programs. The government benefit programs available to eligible citizen include child nutrition programs that provide free school breakfast, lunch, and after school snacks, the Supplemental Nutrition Assistance Program (SNAP), and Temporary Assistance for Needy Families (TANF). From this discussion among experts, modellers need to resolve the discussion into several key elements of the system. One potential set of elements drawn from the expert discussion is shown below:

\begin{itemize}
	\item Government benefits, $B$: the rate at which a particular neighbourhood is participating in all available government programs
	\item Disposable Income, $I$: the average amount of income available for purchasing food in the neighbourhood
	\item Food insecurity, $F$: the rate at which families and individuals experience insufficient access to food
	\item Long-term health outcomes, $H$: measured by an overall health index aggregated at the neighbourhood level.
\end{itemize}

There are several guiding principles to help modellers create a structure that is faithful to the experts' description.

\paragraph{Scope} One common difficulty that appears in many structural elicitation exercises is the tendency of expert groups to think only in terms of measured quantities, rather than underlying drivers. Food insecurity and poverty researchers often consider elements of the system as documented for policy-makers, whereas those with a first hand knowledge of food insecurity may consider a different set of drivers, like personal trauma \cite{Dowler2012, Chilton2009}. Anecdotes of food insecurity may often draw out key, overlooked features of the system, but a well-defined problem scope is critical to prevent a drifting purpose. The responsibility of guiding the conversation continually toward general representations rather than particular instances falls to the modeller. 

\paragraph{Granularity} Elicitations typically begin with a coarse description before beginning to refine the system. Considering refinements and aggregations can help the experts' opinions of the key elements of the system to coalesce. For instance, rather than modelling all the government benefits together in $B$, we could have split this into different variables, child nutrition programs, $C$, and financial support for individuals $S$. The granularity of key elements depends on the modeller's foucs.

\paragraph{Potential interventions} Another guiding principle during the structural elicitation is ensuring that possible interventions are represented by the system components. For instance, if the policy experts wanted to know what would happen after increasing all benefit programs simultaneously, modelling benefits collectively as $B$ would be appropriate. But if they want to study what happens by intervening on child nutrition programs, then separating this node into $C$, child nutrition programs and allowing $B$ to represent additional benefit programs would compose a more suitable model. 

\paragraph{Context Dependence} As the key elements of the system emerge, testing the structure by imagining these key elements in a different structure may either restrict or elucidate additional model features. The drivers that cause food insecurity at the neighbourhood level may vary greatly from those that provoke food insecurity at the individual household level. 

\begin{figure}
	\includegraphics[width=.3\textwidth]{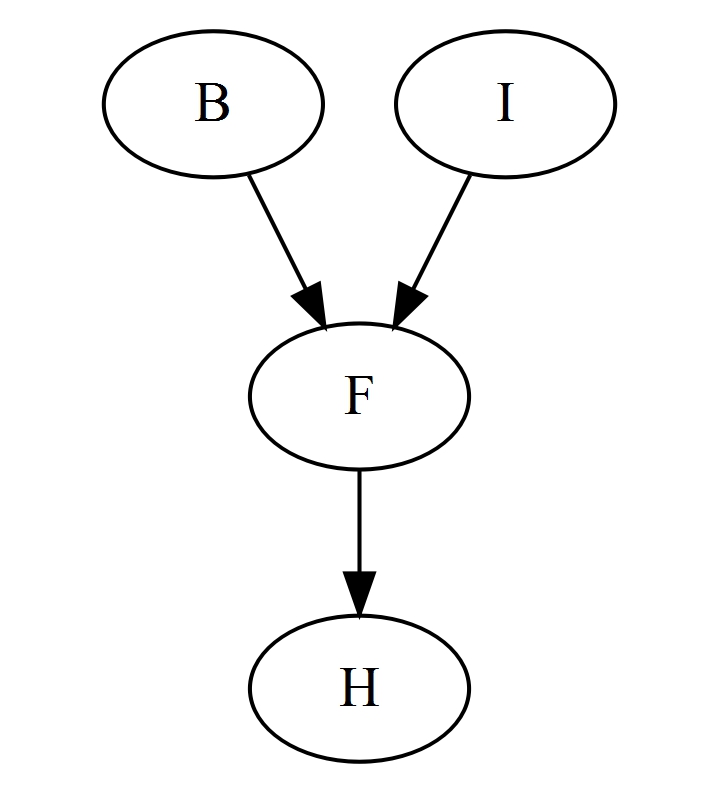}
	\includegraphics[width=.3\textwidth]{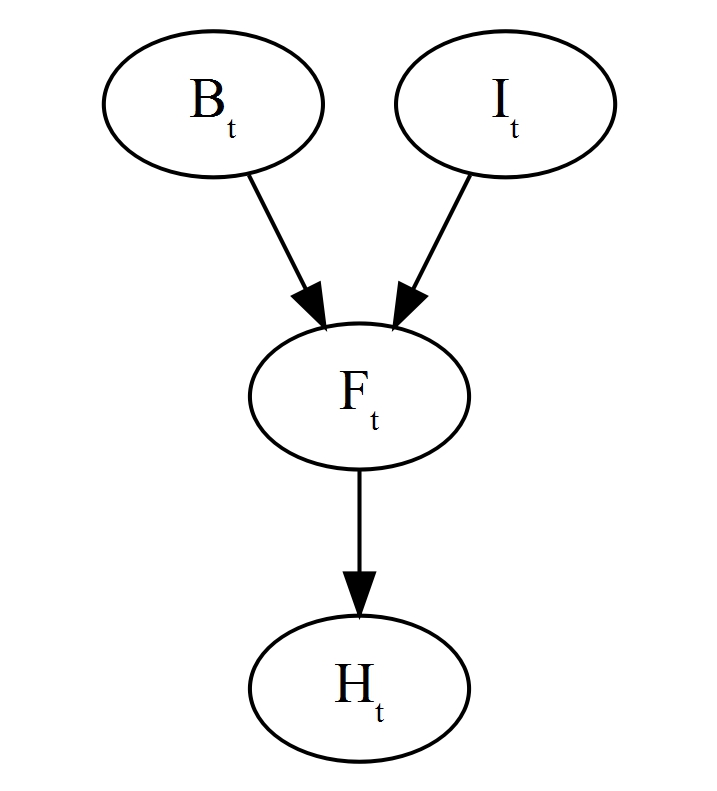}
	\includegraphics[width=.3\textwidth]{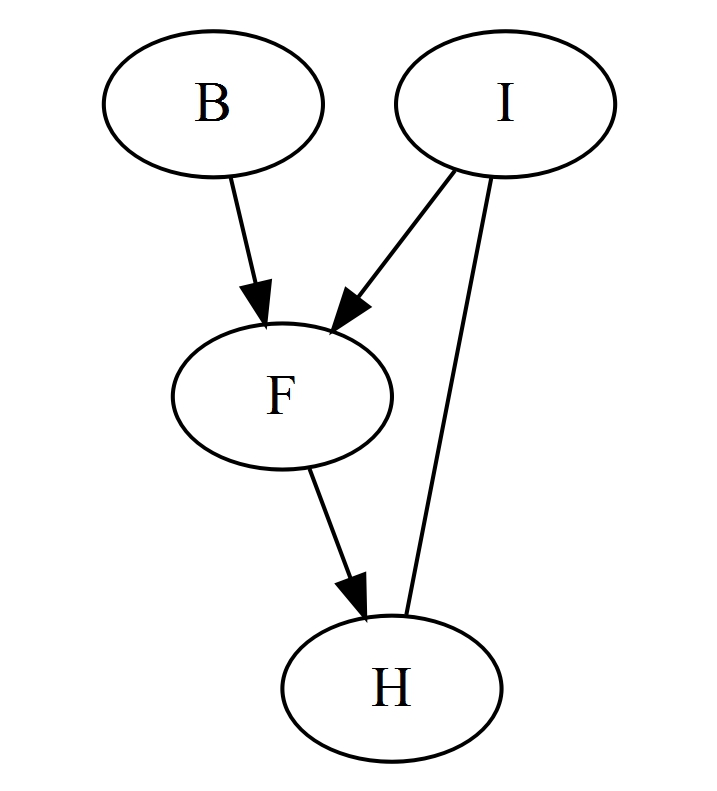}
	\caption{DAG of food insecurity at the neighbourhood level (left). Time series representation of food insecurity drivers over time (centre). Hybrid representation of food insecurity drivers with ambiguous relationship between health and food insecurity (right). }
	\label{fig:BN}
\end{figure}

For this running example, the experts focus on the neighbourhood level. They speak about each of the variables as the particular incidence rates for a neighbourhood.  The modeller could then draw a dependence structure for random variables from their discussion about the dependence between these measurements. This structure would be most conducive to a Bayesian network.
An example of one tentative BN structure that has tried to accommodate these points is on the left of Figure \ref{fig:BN}.

\paragraph{Importance of temporal processes} One key modelling decision in this an other analogous procedures is whether or not to use a dynamic network model. Are the experts speaking about potential interventions that are time-dependent or not? Do the key elements of the process change drastically? Few elements of a system are ever truly static, but dynamic models should only be chosen when the temporal element is crucial to the experts' description of the system.

In contrast to the static example of measurements given above, suppose that the experts believe that yearly fluctuations in disposable income directly affect the rates of food insecurity. This is a dynamic process. Another expert might draw on literature that shows the linear relationship between $I$ and $F$. Using a standard Bayesian network for this problem description would not capture the temporal information or the strength between each of the pairs of nodes. The quantities of the graph here are not static random variables, but rather its nodes appear to be representing processes. In this case, a more appropriate choice for the graphical elements would be to represent them as time series $B_t, I_t, F_t, H_t$. This graph is shown at a single time point in the centre of Figure \ref{fig:BN}. The probabilistic model can be embellished into a number of different stochastic descriptions as will be discussed later in this article.

The meaning of the graph begins to change as the modeller learns more about the structure of a problem. We suggest ways in which we could begin to frame different models for a desired context in terms of nodes and edges. Nodes for general graphical models can be any mathematical objects suitable to the given domain, provided that the system can be actually represented in terms of a probabilistic distribution which is consistent with the meaning we can ascribe to the model edges.  

Once we have established the nodes, the relationships between variables must be represented. These are usually expressed in terms of oriented edges or colourings in the vertices. Continuing with our toy example, the advocates promptly recognize that government benefits and disposable income directly impact the state of food insecurity. It also appears natural, as another expert attests, to associate the long-term health as dependent on food insecurity. These three relationships give us the left graph in Figure \ref{fig:BN}. 

The experts comment that the available money for food purchasing directly affects how much food a family can buy, making directed edges a natural fit for $B$ to $F$ and $I$ to $F$. However, the relationship between long-term health outcomes and disposable income is less clear. One advocate mentions that individuals and families who are battling chronic illness or faced with an outstanding medical bill are less likely to have disposable income, and thus more likely to be food insecure. However, using the typical BN machinery, adding an edge between long-term health outcomes and disposable income would induce a cycle in the graph and thus render the BN inadmissible. 

One common solution would be to simply ignore this information and proceed only with the BN given previously. A second solution would be to embellish the model into a dynamic representation that could formally associate this aspect of the process by expressing instantaneous relationships in a single time slice of effects between nodes on different time slices. A time slice simply denotes the observations of the variables at a given time point.
Another method might be to incorporate an undirected edge that could be used to represent the ambiguous relationship between $I$ and $H$. The result is a hybrid graph with undirected and directed edges with its own logic.

Whatever semantic we choose, edges should represent the experts' natural language description of the relationships. Returning to the instance in which the experts speak about food insecurity as a time series, the edges represent regression coefficients as the system unfolds. As we will show below, directed acyclic graphs (DAGs) are particularly convenient for modelling. However, there are graphical representations that permit cycles, should the modeller wish to focus on the cyclic nature of $F$ and $H$. The choice between the type and orientation of edge affects the semantics of the model as shown below. 

\subsection{Stating irrelevancies and checking conditional independence statements}
\label{sec:condip}
Suppose we choose to represent a client's problem with a BN. Often, it is more natural for experts to impart meaning to the edges present in a graphical model. Unfortunately, it is the absence of edges that represent the conditional independences. To facilitate a transparent elicitation process, these conditional independence relationships can be expressed in a more accessible way as questions about which variables are irrelevant to the other. 

Domain experts who are not statistically trained do not naturally read irrelevance statements from a BN. So it is often important to explicitly unpick each compact irrelevance statement written in the graph and check its plausibility with the domain expert. 

Suppose the domain expert believes that $X$ is irrelevant for predicting $Y$ given the measurement $Z$. That is, knowing the value of $X$ provides no additional information about $Y$ given information about $Z$.  These beliefs can be written as $ X \condip Y \given  Z$, read as $X$ is independent of $Y$ conditional on $Z$. 

Investigating $d$-separation from the graph requires inspecting the moralized ancestral graph of all variables of interest, denoted as $(\mathcal{G}_{\text{An}(A \cup B \cup S)})^m$ \cite{PEARL, JQSdecisionAnalysis}. This includes the nodes and edges of the variables of interest and all their ancestors. Then, we moralize the graph, drawing an undirected edge between all pairs of variables with common children in the ancestral graph. After disorienting the graph (replacing directed edges on the graph with undirected ones) and deleting the given node and its edges, we can check conditional independence between variables of interest.  If there is a path between the variables, then they are dependent in the BN; otherwise they are independent. 

For our example, the missing edges indicates three conditional independence relationships $H \condip B \,|\, F$, $H \condip I  \,|\, F$ and $B \condip I$. to check these, the modeller would ask the following questions to the domain expert: 
\begin{itemize}
	\item If we know what the food insecurity status is, does knowing what the disposable income is provide any additional information about long-term health? 
	\item Assuming we know the food insecurity level, does the government benefit level offer any more insight into the long-term health of a neighbourhood? 
	\item Does knowing disposable income levels of a neighbourhood provide further information about the government benefit level? 
\end{itemize}

This last question might prompt the expert to realize that indeed, disposable income determines eligibility for government benefits, so an edge would be added between $B$ and $I$.

These questions can also be rephrased according to the semigraphoid axioms, rules that simplify properties expect to hold given sets of conditional independence statements. It is helpful to include these as they provide a template for different rule-based styles for other frameworks that capture types of natural language. More details can be found in \cite{JQSdecisionAnalysis}. The first such axiom is given below.

\begin{definition}
	The \textbf{symmetry property} requires that for three disjoint measurements $X$, $Y$, and $Z$: 
	\[
	X \condip Y \given Z \Leftrightarrow Y \condip X \given Z 
	\]
	
\end{definition}

This axiom asserts that assuming $Z$ is known, if $X$ tells us nothing new about $Y$, then knowing $Y$ also provides no information about $X$.

The second, stronger semigraphoid axiom is called perfect composition \cite{Pearl2014}. thus, for any four measurements $X$, $Y$, $Z$, and $W$: 
\begin{definition}
	\textbf{Perfect composition} requires that for any four measurements $X$, $Y$, $Z$, and $W$: \[
	X \condip (Y, Z) \,|\, W \Leftrightarrow  X \condip Y \,|\, (W,Z) \text{ and } X \condip Z \,|\, W 
	\]
\end{definition}

Colloquially, this tells us that assuming $W$ is  known, then if neither $Y$ nor $Z$ provides additional information about $X$, then two statements are equivalent. 
Firstly, if two pieces of information $Y$, and $Z$ do not help us know $X$, then each one on its own also does not help model $X$. 
Secondly, if one of the two is given initially alongside $W$, the remaining piece of information still does not provide any additional information about $X$. Further axioms are recorded and proved in \cite{PEARL}. For the purposes of elicitation, these axioms prompt common language questions which can be posed to a domain expert to validate a graphical structure. 
Given the values of the vector of variables in $Z$, learning the values of $Y$ would not help the prediction of $X$.  Note that when we translate this statement into a predictive model, then this would mean that we know $p(x\,|\,y,z)=p(x\,|\,y)$.

BNs encode collections of irrelevance statements that translate into a collection of conditional independence relationships. This can be thought of as what variable measurements are irrelevant to another. Relationships of the form $X \condip Y \,|\, Z$ can be read straight off the graph as missing edges indicate conditional independence relationships. BNs obey the global Markov property, that each node is independent of its non-descendants given its parents \cite{PEARL}. By identifying the non-descendants and parents of each node, the entire collection of independence relationships is readily apparent. To see this in our example, consider the node representing long-term health, $H$. $\{ B, I\}$ are its non-descendants, and $F$ is its immediate parent, so we know that  $H \condip B  \,|\, F$ and $H \condip I  \,|\, F$. Pearl and Verma \cite{Pearl1991} proved the $d$-separation theorem for BNs, definitively stating the conditional independence queries that can be answered from the topology of the BN in Figure \ref{fig:BN}. The $d$-separation criteria and associated theorems formalize this process of reading off conditional independence relationships from a graph.

\begin{theorem}
	Let $ A, B,  $and $ S $ be disjoint subsets of a DAG $\mathcal{G}$. Then $S$ $d$-separates $A$ from $B$ if and only if $S$ separates $A$ from $B$ in $(\mathcal{G}_{\text{An}(A \cup B \cup S)})^m$, the moralized ancestral graph for the set.
	
	%A set $z$ is said to $d$-separate $x$ from $y$ if and only if $z$ $d$-separates every path form a node in $x$ to a node in $y$. When sets $x$ and $y$ are $d$-separated by $z$ then $x$ is independent of $y$ conditional on $z$ in every distribution compatible with the graph. 
	
\end{theorem}
Proof given in \cite{Lauritzen1996}. 

As an example, consider a BN of the drivers of food insecurity shown in Figure \ref{fig:BN}. 
The $d$-Separation theorem tells us that $H$ is $d$-separated from $B$ and $I$ given $F$. Graphically, $d$-separation can be investigated by examining the moralized ancestral graph. A moralized graph is one in which there is an undirected edge between the parents of a node. We see that in the moralized graph, $F$ $d$-separates every path from the node $H$ to a node in the set $\{B, I\}$. Thus, $d$-separation allows us to consider the relationships between any three subsets of variables in the DAG.

Separation theorems have been found for more general classes of graphs including chain graphs, ancestral graphs, and chain event graphs \cite{Studeny1995, Andersson2001, Richardson2002}. Another class of graphical model, vines, weakens the notion of conditional independence to allow for additional forms of dependence structure \cite{Bedford2002}. 
The results of the separation theorem for BNs can also be used to explore independence relationships in classes of graphs that are BNs with additional restrictions such as those imposed by the Multi-regression Dynamic models \cite{Smith1993} and flow graphs \cite{Smith2007}.

When the structure is verified, it can then be embellished to a full probability model, provided it meets the original assumptions of our model. 
Understanding the relationship between the elicited conditional independence statements implied by the graph ensures that we do not elicit equivalent statements, thereby reducing the number of elicitation tasks. Even more importantly, the probabilities will respect the client's structural hypotheses--hypotheses that are typically much more securely held than their numerical probability assessment.

In a discrete BN, this process involves populating the conditional independence tables with probabilities either elicited from experts or estimated from data. Our food insecurity drivers example could be embellished to a full probability representation of a continuous BN. In this case, the full probability distribution $p(x)$ would be given by the factorization of the joint distribution: 

\[
p(x)=p(i)p(b \,|\, i) p(f \,|\, i, b) p(h \,|\, f) 
\]

These distributions can be either discrete or continuous. Discrete BNs will be populated by conditional independence tables that assign probabilities to all possible combinations of the values of each term in the factorized joint probability density. In continuous BNs, hybrid methods sample a continuous BN once, and then discretize each of the nodes to update the BN quickly \cite{Hanea2007}. This allows for scalable inference and updating of the BN in a high-dimensional, multivariate setting. The probabilities underpinning this model can be elicited using additional protocols and procedures from other chapters of this book.

\section{Examples from food insecurity policy}
\label{sec:foodex}

\subsection{Bayesian network}
\label{sec:bn}

Structural elicitation for a Bayesian network is well studied \cite{JQSdecisionAnalysis, Korb2009}. To see this process in action, consider a food insecurity example. The United States Department of Agriculture (USDA) administers the national School Breakfast Program (SBP), serving free or reduced price meals to eligible students. 

A key element of the system is understanding the programmatic operations. Participation in SBP is not as high as it is for the school lunch program \cite{Nolen2015}. The traditional model of breakfast service involves students eating in the cafeteria before the beginning of school. Advocates began promoting alternative models of service to increase school breakfast participation. These include: Grab n Go, in which carts are placed through the school hallways and students select a breakfast item en route to class, or Breakfast in the Classroom, where all students eat together during the first period of the day. 
Only schools which have 80\% of students eligible for free or reduced lunch are eligible for universal school breakfast. This means that breakfast is offered to every child in the school, regardless of their free or reduced status. This policy  was implemented to reduce stigma of receiving a free meal. 

The experts would also like to understand the effects of not eating breakfast. Advocates, principals, and teachers have hypothesized that eating school breakfast impacts scholastic achievement. Food-insecure children struggle to focus on their studies. Schools also show a reduced rate in absenteeism, as children and parents have the added incentive of breakfast to arrive at school. Some evidence suggests eating breakfast may also reduce disciplinary referrals, as hungry children are more likely to misbehave.

The data for this problem comes from a set of schools who are all eligible for universal breakfast, but some have chosen not to implement the program while others have. As universal breakfast status can be used as a proxy for socio-economic background of students attending a school, we have narrowed the population to schools with low socio-economic status. The group of experts do not describe a temporal process here. They do not mention changes in breakfast participation throughout the school year, yearly fluctuations, or a time series of participation rates. Thus, it is natural for the modeller to begin with a BN approach. Given this information about breakfast, led by a facilitator, the modeller could consolidate the discussion into the following nodes:

\begin{itemize}
	\item $X_1$ Model of Service (Yes, No): indicates whether or not an alternative model of service as been implemented 
	\item $X_2$ Universal (Yes, No): indicates whether or not an eligible school has opted into universal service, as opposed to checking the economic status of the student at each meal
	\item $X_3$ Breakfast Participation (High, Medium, Low): the binned participation rates at each school
	\item $X_4$ Scholastic Achievement (High, Medium, Low): the standardized test score for each school
	\item $X_5$ Absenteeism (High, Low): the binned absenteeism rate for the year
	%\item $X_ 6$ Disciplinary Referrals (High, Low): absolute number of disciplinary referrals %how do the binned vs absolute numbers change things in a BN???
\end{itemize}

We note that this list of nodes is focused on understanding the effects of school breakfast participation and specific type of breakfast service model. 
Certainly there are other reasons for absenteeism and disciplinary referrals besides whether or not a student had a good breakfast, but these are beyond the scope of this model. How do we determine the structure of this model from these measurable random variables? 
From this set of nodes, the decision maker is queried about the possible relationship between all possible sets of edges. For instance, we could ask, does knowing whether or not the school has opted into universal breakfast give any other information about whether or not the school has implemented an alternative breakfast model? 
In this case, the decision makers believe $X_1$ does not give any additional information about $X_2$, because the program model is subject to approval from the cafeteria managers and teachers, whereas the decision to implement universal breakfast is primarily the decision of the principal. 
Thus no edge is placed between $X_1$ and $X_2$. Both $X_1$ and $X_2$ are helpful in predicting $X_3$, so an arrow is drawn between each of these pairs. $X_4$ is affected by $X_3$. 

It is important to note that if we had taken the population to be all schools rather than those with a low socio-economic status, then $X_2$ would affect $X_4$, $X_5$, and $X_6$ because universal school lunch would then be a proxy for low socio-economic status.

\begin{figure}[htp]
	
	\centering
	\includegraphics[width=.3\textwidth]{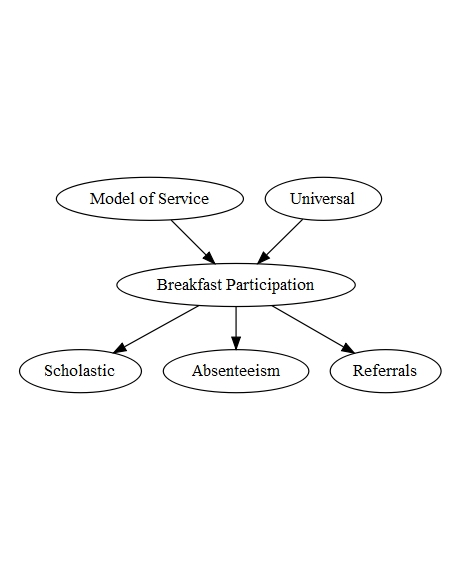}\hfill
	\includegraphics[width=.3\textwidth]{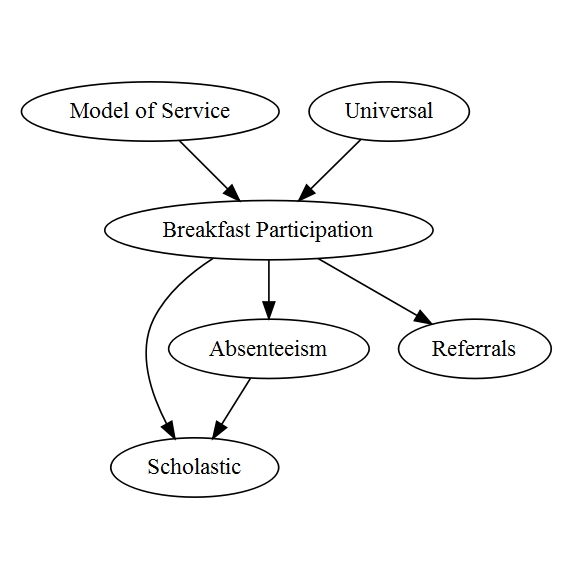}\hfill
	\includegraphics[width=.3\textwidth]{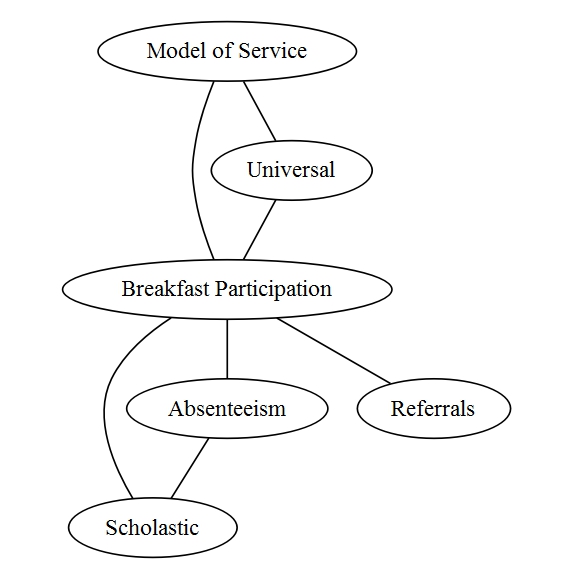}
	
	\caption{The figure on the left is the original BN representing the effects of model service on breakfast participation and academic outcomes. The central BN represents the original BN with an edge added through the described verification process. The right graph represents the ancestral, moralized DAG of the central BN.}
	\label{fig:breakfastBN}
	
\end{figure}

Suppose we know a school has a low breakfast rate, and we want information about their absenteeism. Will knowing anything about scholastic achievement provide any additional information about absenteeism? In order to check this with $d$-separation, we examine the ancestral graph $\mathcal{G}_{\text{An}(X_4, X_5, X_6)}$, the moralised graph $(\mathcal{G}_{\text{An}(X_4, X_5, X_6)})^m$. If there is not a path between $X_4$ and $X_5$, then we can say that $X_4$ is irrelevant to $X_5$. However, if there is a path between $X_4$ and $X_5$ that does not pass through our given, $X_3$, then the two variables are likely to be dependent. Thus, we can use the $d$-separation theorem to check the validity of the BN. We may also ask equivalent questions by symmetry. For instance, suppose we know a school has a low breakfast rate and we want to know information about their scholastic achievement. Will additional information about absenteeism be relevant to scholastic achievement? Asking such a question may prompt our group of decision makers to consider that students who miss classes often perform worse on exams. Revising the BN is in order, so we add an additional edge from $X_5$ to $X_4$. The BN in Figure \ref{fig:breakfastBN} represents the beliefs of the domain experts. this encodes the following irrelevance statements:

\begin{itemize}
	\item Knowing the model of service provides no additional information about whether or not the school district has implemented universal breakfast.
	\item The model of service provides no additional information about scholastic achievement, absenteeism, or referrals given that we know what the percentage of students who eat breakfast is. 
	\item Knowing absenteeism rates provides no additional information about disciplinary referrals given that we know what the breakfast participation rate is.
	\item Knowing scholastic achievement rates provides no additional information about disciplinary referrals given that we know what the breakfast participation and absentee rates are.
\end{itemize}

When these irrelevance statements are checked, the domain experts realize that there is an additional link in that absenteeism affects scholastic achievements. Thus we draw an additional arrow between $X_4$ and $X_5$. The relationship between referrals and absenteeism is disputed in the literature, so, at least in this first instance, we omit this edge.  

Once the experts agree on the structure and verify it using the irrelevance statements, then the modeller may elicit the conditional probabilities. Taken together, the BN represent a series of local judgements. The joint density of a BN is represented by:
\[
p(\bm{x}) = \prod_{i=1}^n p(x_i | \text{pa}(x_i))
\]

For our example,  
\[
p(\bm{x}) = p(x_1)p(x_2), p(x_3 | x_1,x_2), p(x_4 | x_3, x_5), p(x_5 | x_3) p(x_6 | x_3)
\]

Many of these probabilities may be estimated by data, and unknown quantities may be supplied through structured expert elicitation. 
For instance, consider the sample question: what is the probability that scholastic achievement is high given that breakfast participation rate is medium and the absentee rate is low?  
When the conditional probability tables are completed, the BN can be used to estimate effects of intervention in the system according to \cite{PEARL}.

\subsection{Chain Event Graph}
\label{sec:ceg}

To illustrate an instance when a bespoke representation is more appropriate than the BN, consider the example of obtaining public benefits to address food insecurity. The USDA's Supplemental Nutrition Assistance Program (SNAP) provides funds for food to qualifying families and individuals through Electronic Benefit Transfer (EBT). Although 10.3\% percentage of Americans qualify for the program. Loveless \cite{Loveless2012} estimates that many more citizens are eligible for benefits than actually receive them. Policy makers and advocates want to understand what systematic barriers might prevent eligible people from accessing SNAP. The application process requires deciding to apply, having sufficient documentation to apply (proof of citizenship, a permanent address), a face to face interview, and correct processing of the application to receive funds.

The structural elicitation phase includes speaking with domain experts to gather a reasonably comprehensive list of steps in the process. Domain experts include case workers, advocates, and individuals applying through the system. For our example, Kaye et al. \cite{Kaye2013} collected this information through interviews at 73 community based organizations in New York State and categorized it according to access, eligibility, and benefit barriers. This qualitative information collection is crucial to developing an accurate model. 
From the qualitative studies, the key barriers were identified as: 

\begin{itemize}
	\item Face to face interviews not waived
	\item Same-day application not accepted
	\item Excessive documentation required
	\item Expedited benefit (available to households in emergency situations) not issued
	\item Failed to receive assistance with application documents
	\item Barriers experienced by special population: elderly and immigrant
	\item Ongoing food stamp not issued within 30 days
	\item EBT functionality issues
\end{itemize}

The events selected should be granular enough to encompass the key points at which an applicant would drop out of the process, but coarse enough to minimize model complexity. 
An important part of the qualitative analysis process includes combining anecdotal evidence into similar groupings. 
For instance, the benefits office refused to waive the in-office interview for an applicant who did not have transportation to the application centre. 
In a separate instance, an interview was not waived for a working single mother with four children who could not attend because she was at work. While there are different contexts to each example, the central problem is the failure to waive the face to face interview. 
This type of node consolidation aids in reducing model complexity.  

Discretizing events can be a convenient, way to clarify the model structure. Checking that the discretisation covers all possible outcomes from that event ensures that the model is an accurate representation of the problem. For our example, one possible discretization with four variables of the problem is: 

\begin{itemize}
	\item $x_r$: At-risk population? (Regular, Elderly, Immigrant)
	\begin{itemize}
		\item Regular: Households not part of an at-risk population
		\item Elderly: Household head is over 65
		\item Immigrant: Household head is a citizen, but immigration status of members of the household is uncertain
	\end{itemize} 
	\item $x_a$: Decision to apply (Expedited, Regular application, Decides not to apply) 
	\begin{itemize}
		\item Expedited: Same day applications, used in cases of emergency food insecurity
		\item Regular application: The standard procedure
		\item Decides not to apply: Eligible households who elect not to apply for a variety of reasons
	\end{itemize}
	\item $x_v$: Application Verdict (Rejected, Accepted, Revision Required)
	\begin{itemize}
		\item Rejected: Failed application, no possibility of resubmission
		\item Accepted: Successful application
		\item Revision required: Application must be resubmitted because of missing documentation, missed interview, or other reasons
	\end{itemize}
	\item $x_e$: Utilizing an EBT card (Card successfully used for transactions, transaction errors)
	\begin{itemize}
		\item Card used for transactions: EBT arrives within the 30 month deadline and is successfully  used at a grocery store
		\item Transaction errors: Card either does not arrive or returns an error at the grocery store
	\end{itemize}
\end{itemize}

\begin{figure}[h]
	
	\centering
	\includegraphics[width=.7\textwidth]{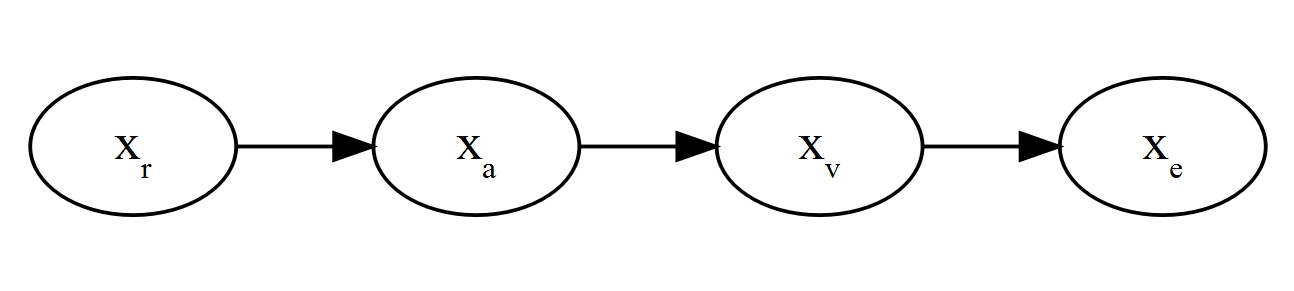}
	\caption{An inadmissible BN for the public benefits application process example.}
	\label{fig:snapBN}
\end{figure}

Figure \ref{fig:snapBN} shows a simple BN approach to the natural language problem. Assume that the conditional independence relationships have been checked and that we can now supply the conditional probabilities. As we begin this process, note that some of the probabilities are nonsensical. For example, we must supply a probability for quantities like: the probability of having an accepted application given that the eligible citizen decided not to apply, and the probability of successfully utilizing EBT given that the application was rejected. This probability setting sounds absurd to elicit, and will be distracting during the probability elicitation.

%sidecaption 

\begin{figure}
	
	\centering
	\includegraphics[width=.5\textwidth]{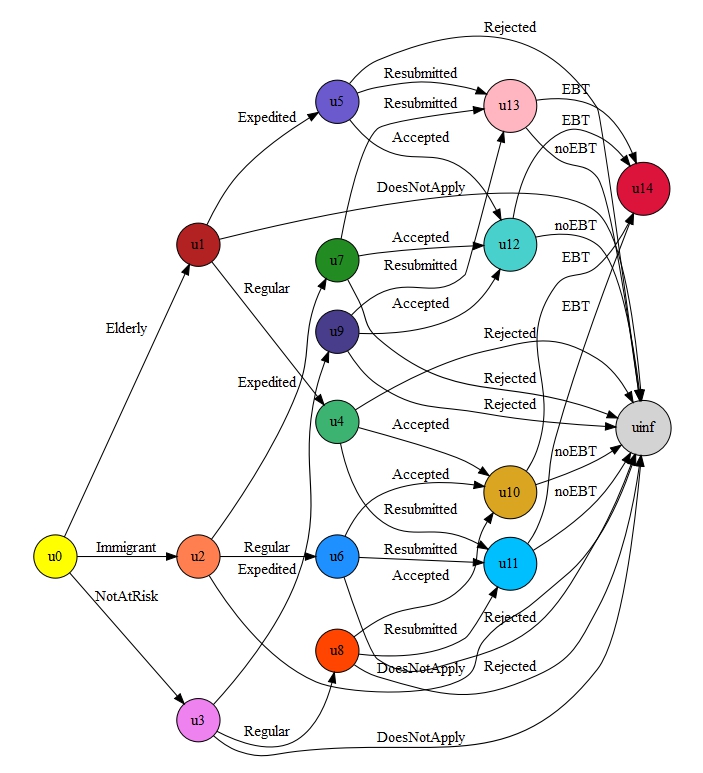}
	\caption{Chain Event Graph representation of the benefits application process.}
	\label{fig:ceg}
	
\end{figure}

The application process is difficult to coerce into a BN because the problem is highly asymmetrical For instance, applicants with insufficient documentation will not have the chance to interview, and will not progress through the system. Now, if we consider again the natural language of the experts, we notice that this process is described as a series of events that have a natural ordering. Applicants must first decided to apply, then receive a verdict, and finally use their EBT card. The notion of being a member of an at-risk population does not have an explicit ordering, but we can reasonably order it before the other events as it may affect how downstream events unfold. Collazo et al. \cite{CEGbook} show that ordering demographic information at the beginning often coincides with higher scoring models during model selection for this class of graphs.  Shafer \cite{Shafer1996} has argued that event trees are a more natural way to express probabilistic quantities, so we will instead express this problem as an event tree on the left of Figure \ref{fig:ceg}. We next show how, in this instance, there is an alternative graphical framework that provides a better way of accommodating the information provided by the expert.

The nodes of our event tree are called positions; they represent different situations of applicants travelling through the system. The edges represent the different outcomes of each possible event. We can elicit the probability of observing a unit travelling down each edge of the tree. The probability of a unit travelling down each of those edges should sum to one for each position. The root to sink paths on the tree can be thought of as all possible outcomes of the application procedure. Situations with the same colour on the tree represent positions whose outgoing edges have the same probabilities. In Figure \ref{fig:ceg}, leaf nodes showing terminating outcomes are depicted in light grey. 

The tree structure is naturally flexible just like the BN and can easily be modified to accommodate natural language suggestions. 
For instance, suppose the expert would like to add in a variable: the outcome of an interview process for regular applicants (the expedited process is waived.) This can easily represent diversity onto the tree structure, whereas it would require adding entries to the conditional probability tables for more nodes. 

Another feature of the event tree structure is that the context specific independences are expressed directly in the tree structure. 
Elderly applicants are often less likely to apply for benefits because the dollar amount is often too small a motivation for the perceived difficulty of the application. Immigrants are also less likely to apply because, although citizenship is required to apply for benefits, citizens with undocumented family members may fear citizenship repercussions of applying for assistance. 
These context-specific probabilities are modelled through the colourings of the nodes and vertices of the CEG, rather than requiring separate BN models with context-specific conditional independence relationships. 

\begin{figure}[h]
	
	\centering
	\includegraphics[width=.7\textwidth]{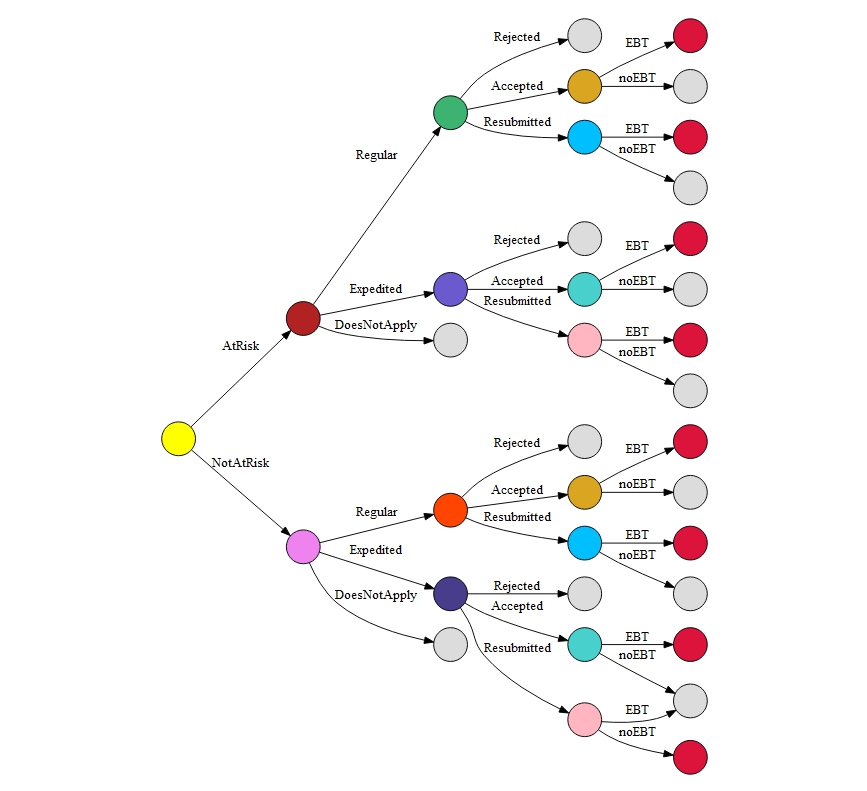}
	\caption{Event tree depicting the outcomes of the benefit application process.}
\end{figure}

The same elicited probabilities with equivalent sub-trees represent stages that can be merged for a more compact chain event graph representation, called the Chain Event Graph (CEG), depicted on the right of Figure \ref{fig:ceg}.
Next, we validate the structure of the CEG using the $d$-separation theorem and equivalent questions using symmetry in a way analogous to the BN theorem. Thwaites et al. \cite{Thwaites2015} discovered a new $d$-separation theorem for CEGs. Given that a unit reaches a position, what happens afterwards is independent not only of all developments through which it was reached, but also of the positions that logically can't happen. These conditional independence statements can be read off the graph just as they can for BNs. We illustrate this process below.

Let $\mathcal{T} = (V,E)$ denote a staged tree, a tree coloured to represent conditional independence. Then, we call $W \subseteq V$ a fine cut if disjoint union of events centred on these vertices is the whole set of root-to-leaf paths. 
When this is the case, none of the vertices $w \in W$ are up or down stream of the other, and the union of paths including $w \in W$ is the entire tree. 

Furthermore, $W' \subseteq U_{\mathcal{T}}$ is a cut if the set $\{ v \in u | u \in W'\}$ is a fine cut. 
Denote $y_{\prec W} = (y_w | w \text{ upstream of } W)$ and $y_{W \preceq} = (y_{w'} | w' \text{ downstream of } W)$. Then for any cut-variable $x_{W'}$,
$x_W$ is a cut variable if $W$ is a cut  or a fine cut and $x_W$ is measurable with respect to the probability space defined by the underlying stage tree. 

\begin{theorem}
	Let $\mathcal{C} = (W,E)$ be a CEG and let $W' \subseteq W$ be a set of vertices then for any cut-variable $x_{W'}$, we find:
	\begin{enumerate}
		\item If $W'$ is a fine cut then $y_{\prec W'} \condip y_{W' \preceq} | x_{W'}$. 
		\item If $W'$ is a cut then $y_{\prec W'} \condip y_{W'} | x_{W'}$. 
	\end{enumerate}
\end{theorem}

Proof can be found in \cite{SmithAnderson2008}.

The CEG can express conditional independence between functions of random variables rather than subsets of a given vector of random variables. %TODO HUH??
For each cut of variables, natural language questions from the semigraphoid axioms elucidate the conditional independence relationships. At each cut, consider the conditional independence between each pair of upstream and downstream variables.  For instance, given that eligible applicants apply for benefits, does knowing whether or not they are part of an at-risk population provide any additional information about whether or not they apply for expedited benefits? By perfect decomposition, does knowing that the candidate received application assistance provide any information about whether or not they will receive the electronic benefits given that they had the correct documentation and passed the interview? Does knowing that they had application assistance provide any additional information about whether or not they passed the interview given that they had the correct documentation?  
These queries validate the model and may prompt further adaptations. 

In a BN, the ancestral graph helps to address these queries. The ancestral graph has no direct analogue in the CEG. Instead, following \cite{Thwaites2015} the CEG admits a pseudo-ancestral representation. Pseudo-ancestral graphs depict the nodes of interest and all the upstream variables, consolidating the downstream variables. Moralizing the graph in a BN corresponds to removing the colourings of the CEG. 

\begin{figure}[h]

	\includegraphics[width=.5\textwidth]{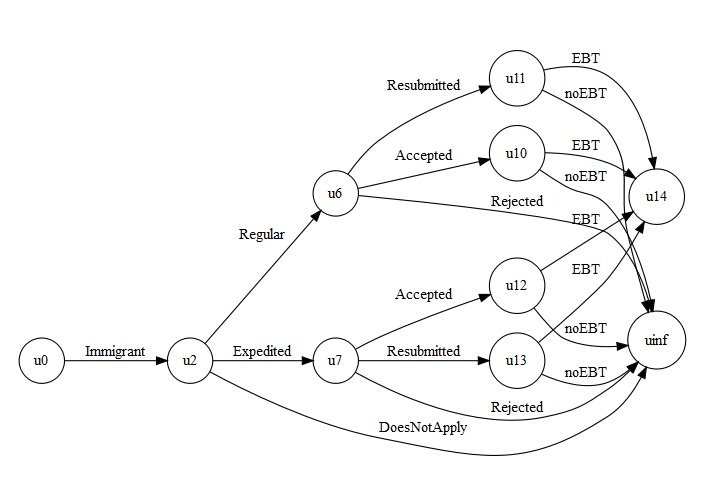}
	\includegraphics[width=.5\textwidth]{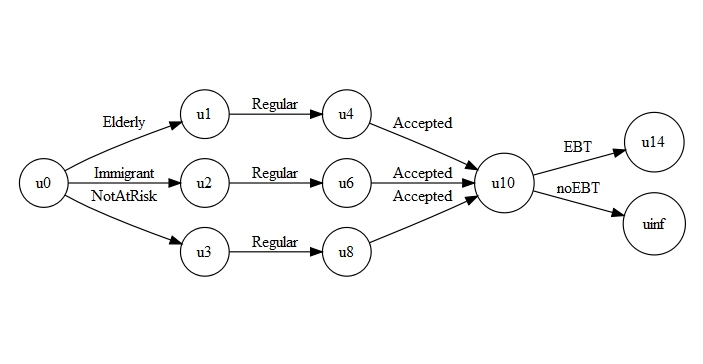}
	\caption{Two uncoloured pseudo-ancestral CEGs}
	\label{fig:ceg_ances}
\end{figure}

Is the ability to complete a transaction on the EBT card independent of whether the applicant is a member of an at-risk population given that they completed a successful regular application? The pseudo-ancestral graph as seen on the right in Figure \ref{fig:ceg_ances}, shows the probability that $\Lambda= \{ \text{Regular, Accepted} \}$. Being a part of the at-risk population is independent of being able to utilize an EBT card because we see that all the possible pathways must pass through $u_{10}$, identifying it as a cut-vertex. 

On the other hand, suppose we want to test the independence of the application verdict from the selected method of application for at-risk immigrant population. This ancestral graph can be given by the left CEG in Figure \ref{fig:ceg_ances}. These are not independent because there is no cut vertex.

One of the strengths of the CEG model is that it does not require any algebra, but instead can be elicited entirely using coloured pictures. 
After validating the structure, populating the model with data or elicited probabilities provides a full model that can be used for inference, details can be found in \cite{CEGbook}. The CEG offers a class of models that is more general than BNs, enabling modellers to represent context-specific independences.

The CEG is a powerful model particularly well-suited to expert elicitation, as experts often convey information in a story, which naturally expands to an event tree. 

\subsection{Multi-regression Dynamic Model}
\label{sec:mdm}

Our next two examples of customised classes of graphical models consider the problem of assessing participation in the Summer Meals Program (SMP).  SMP meal sites are designated as either open or closed. Open sites do not have a set population like in a school or particular program, but rather are open to the public and thus dependent on walk-ins for the bulk of participation. 

Although the need in the summer is severe, participation in the program remains relatively low. Advocates generally agree that the two biggest obstacles to program participation are a lack of awareness about the program, and unavailable transportation to the site. These factors affect meal participation which fluctuates throughout the three months of summer holidays. Available data for meal participation records how many meals were served through the program at each day for about three moths in the summer. Transportation data records the number of available buses. Awareness can be measured through texting data that records when participants queried a government information line to receive information about where the closest sites serving meals are.  

Advocates would most like to capture the effect that awareness of SMP has on available transportation, and that transportation in turn has on meal participation. To simplify the elicitation, additional obstacles like low summer school enrolment, poor food quality, and insufficient recreational actives are not considered as primary drivers of meal participation levels. The relationship between awareness and available transportation is well documented, as is the relationship between transportation and meal participation \cite{Wilkerson2015}. 

The advocates emphasize drastic shifts in awareness, transportation, and meal participation throughout the summer months. On public holidays and weekends, there is a lack of public transportation and a corresponding sharp decline in meals. This temporal aspect of the problem prompts the modeller to consider a time series representation as the most natural class of graphical model.

\begin{figure}
	
	\centering
	\includegraphics[width=.45\textwidth]{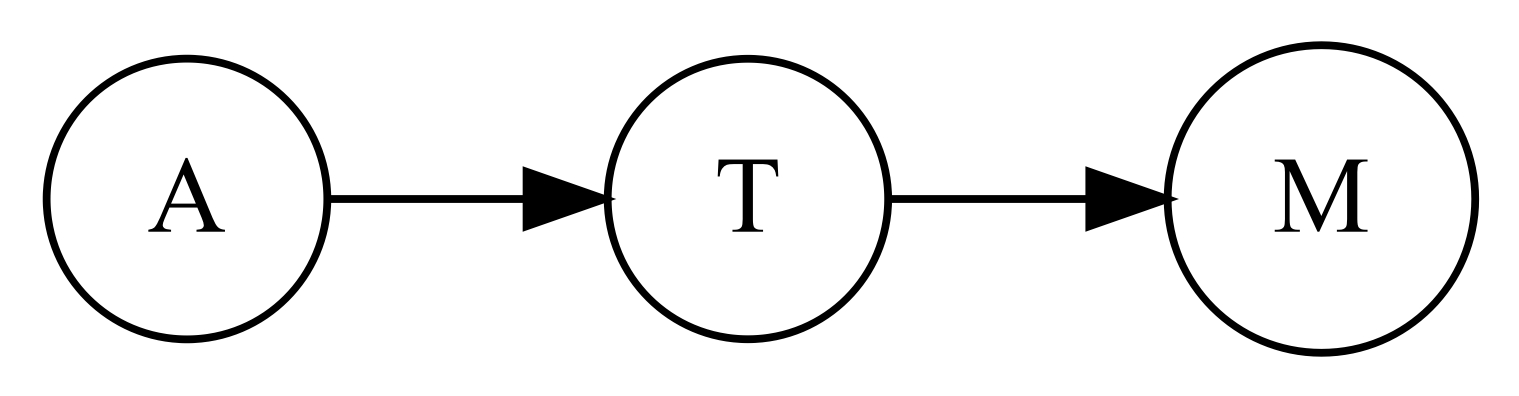} \qquad
	\includegraphics[width=.45\textwidth]{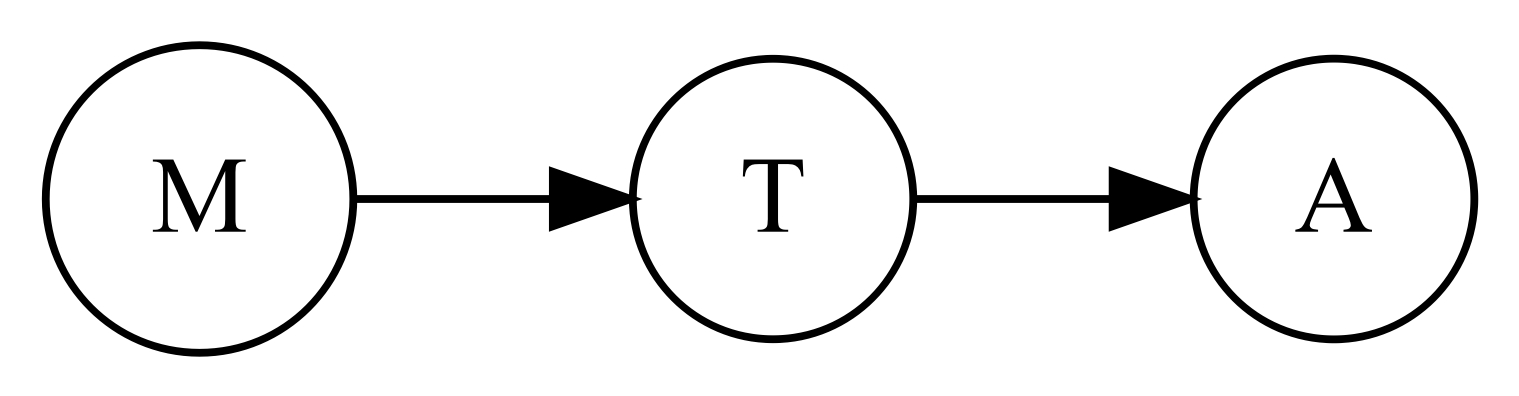}
	\caption{Two DAGs with equivalent BN representations, but unique MDM representations}
	\label{fig:mdm}
\end{figure}

To emphasize the importance of selecting a time series representation over a BN, consider the limitations of the standard BN model. Suppose the advocates agree on the BN structure shown in the DAG on the left of Figure \ref{fig:mdm}, as children and parents must know about the meal before they take transportation to the meal. Then in turn, they must travel to the meal before receiving the meal. However, the graph in Figure \ref{fig:mdm} (L) encodes conditional independence relationship encoded in this representation, $m \condip a \,|\, t$, which does not capture the ordering expressed by the advocates. To further stress this point, Figure \ref{fig:mdm} (R) shows a BN with the reverse ordering that encodes equivalent conditional independence relationships. 

The experts remark that a media campaign and corresponding surge in awareness prompts a corresponding increase in the number of people travelling to meal sites. These aspects of the problem, taken with those discussed above prompt us to consider each of the elements as time series. In order to capture the linear relationship between variables that the experts have expressed, we also define the edges of the graph to correspond to regression coefficients between each parent and child.

Assuming linear relationships exist between awareness and transportation and transportation to the meal site and actual participation, the system can be described as regressions in a time series vector $\bm{Y}_t =\{A_t, T_t, M_t\}$. We denote the time series of the key measurements: awareness by $A_t$, available transportation by $T_t$, and summer meals participation by $M_t$. This model corresponds to another example from our toolbox of alternative representations: the Multi-regression Dynamic Model, the general definition of which is shown below.

\begin{definition}
	A collection of time series $ \bm{Y}_t = \{ Y_t(1), \ldots, Y_t(i), \ldots, Y_t(n)  \}$ can be considered a the Multi-regression Dynamic model (MDM) if three observation equations, a system equation, and initial information as given below adequately describe the system. Each series in the MDM can be represented by an observation equation of the form: 
	
	\[
	Y_t(r) = \bm{F}_t(r)' \bm{\theta}_t(r) + v_t(r) \qquad v_t(r) \sim (0, V_t(r)), \,\, 1 \leq r \leq n
	\]
	with system equation: 
	
	\[
	\bm{\theta}_t = G_{t}\bm{\theta}_{t-1} + \bm{w}_t \qquad \bm{w}_t \sim (0,W_t);
	\]
	and initial information: 
	\[
	(\bm{\theta}_0 | D_0) \sim (\bm{m}_0, C_0).
	\]
	
\end{definition}

$\bm{F}_t(r)$ is a known function of $\bm{y}_t(r)$ for $1 \leq r$. that is, each observation equation only depends on the past and current observations rather than the future ones.
$V_t(r)$ are known scalar variance observations. these can be estimated from available data or else elicited from experts. For our example, $V_t(a) = $, $V_t(t) = $, and $V_t(m) = $. 
$\bm{W}_t = \text{blockdiag}\{W_t(1), \ldots, W_t(n)\}$ and $C_0 = \text{blockdiag}\{C_0(1), \ldots, C_0(n)\}$
are assumed to be known and functions of the past variables of the upstream variables.  The errors must also be independent. This means that  $(\bm{Y}_t(r) | \bm{Y}^{t-1}, \bm{F}_t(r),\bm{\theta}_t(r))$ follows some distribution with mean $\bm{F}_t(r)^t\bm{\theta}_t(r)$ and variance $V_t(r)$.

Particular observations of awareness, transportation, and meal participation are denoted as $\{a_t, t_t, m_t\}$. Modelling this behaviour requires dynamic linear models in which the parents given the regression coefficients for each parent node. For our example, the system and observation model equations are: 

\begin{align}
\bm{\theta}_t(a) = \bm{\theta}_{t-1}(a) + w_t(a) &\qquad a_t = \theta_{t}^{(1)}(a) + v_t(a)\\
\bm{\theta}_t(t) = \bm{\theta}_{t-1}(t) + w_t(t) &\qquad t_t= \theta_{t}^{(1)}(t)  + \theta_t^{(2)}(t) a_t + v_t(t)  \\
\bm{\theta}_t(m) = \bm{\theta}_{t-1}(m) + w_t(m) &\qquad m_t = \theta_{t}^{(1)}t(m)   + \theta_{t}^{(2)} t_t + v_t(m)\\
\end{align}

The initial information $\{\bm{\theta}_0 \}$ can be elicited from the domain experts. 

Suppose after the experts agree on the structure, the modeller examines the one step ahead forecasts, and notices errors on some days. Examining these days might prompt the experts to recognize that the days of interest correspond to days with a heat advisory. They suggest that the heat index throughout the summer $h_t$ also affects meal participation, $m_t$. This structural change can be quickly integrated into the system by adding observation and system equations and initial information for $h_t$ and updating the system for downstream node:

\begin{align}
\bm{\theta}_t(h) = \bm{\theta}_{t-1}(h) + w_t(h) \\
h_t = \theta_{t}^{(1)}t(h)  + v_t(h)\\
m_t = \theta_{t}^{(1)}t(m)   + \theta_{t}^{(3)} t_t + \theta_{t}^{(2)} h_t + v_t(m)
\end{align}  

In this way, the natural language expressions of the domain experts can be used to adjust the model.
the MDM ensures two critical conditional independence relationship. the first holds that if

\begin{equation}
\condip_{r=1}^n \bm{\theta}_{t-1}(r) | \bm{y}^{t-1}
\end{equation}

then 
\begin{equation} \label{eq:mdmcondip1}
\condip_{r=1}^n \bm{\theta}_{t}(r) | \bm{y}^{t}
\end{equation}

and 
\begin{equation}\label{eq:mdmcondip2}
\bm{\theta}_t(r) \condip Y^t(r+1),\ldots, Y^t(n) | Y^t(1), \ldots, Y^t(r) 
\end{equation}

Equation (\ref{eq:mdmcondip1}) tells us that if the parameters$\{\bm{\theta}_{t-1}(r)\}$ are independent of each other given the past data $\{\bm{y}^{t-1}\}$ then $\{\bm{\theta}_{t}(r)\}$ is also independent of $\{\bm{y}^{t}\}$. By induction, we can see that given the initial parameters $\{\bm{\theta}_{0}(r)\}$ are independent, then they remain independent as the series unfolds.

For our example, we need to ensure that $\bm{\theta}_0(a) \condip \bm{\theta}_0(t) \condip \bm{\theta}_0(m)$. Awareness is measured by the amount of public media generated, transportation is a measure of public transportation available, and the participation rate is the number of meals served every day in the summer. The domain experts agree that these can be independent of each other. 
Additionally, equation \ref{eq:mdmcondip2} ensures the following conditional independence relationships:
\begin{align}
\bm{\theta}_t(a) & \condip \{t^{t-1},m^{t-1}\} | a^{t-1}\\
\bm{\theta}_t(t) & \condip m^{t-1} | \{a^{t-1},t^{t-1}\}
\end{align}

An analogue of the $d$-separation theorem for MDMs identifies part of the topology of the graph that ensures that these conditional independence statements hold.
\begin{theorem}
	For MDM $\{\bm{Y}_t\}$ if the ancestral set $\bm{x}_t(r) = \{y^t(1), \ldots, y^t(r)\}$ $d$-separates $\bm{\theta}_t(r) $ from subsequent observations $\{y^t(r+1),\ldots, y^t(n)\}$ for all $t \in T$, then the one-step ahead forecast holds :
	\[
	p(\bm{y}_t | \bm{y}^{t-1}) = \prod_{r} \int_{\bm{\theta}_t(r)} p\{\bm{y}_t(r)|\bm{x}^t(r),\bm{y}^{t-1}(r),\bm{\theta}_t(r)\} \,p\{\bm{\theta}_t(r) |\bm{x}^{t-1}(r),\bm{y}^{t-1}(r)\} d\bm{\theta}_t
	\]
\end{theorem}

\begin{figure}
	\centering
	\scalebox{.75}{
	\begin{tikzpicture}[x=1pt,y=1pt]
	\definecolor{fillColor}{RGB}{255,255,255}
	\path[use as bounding box,fill=fillColor,fill opacity=0.00] (0,0) rectangle (505.89,361.35);
	\begin{scope}
	\path[clip] ( 49.20, 61.20) rectangle (480.69,312.15);
	\definecolor{drawColor}{RGB}{46,139,87}
	
	\path[draw=drawColor,line width= 0.4pt,line join=round,line cap=round] ( 65.18,257.25) --
	( 68.78,264.86) --
	( 72.38,206.35) --
	( 75.98,160.82) --
	( 79.58,278.15) --
	( 83.18,294.64) --
	( 86.78,299.37) --
	( 90.38,301.81) --
	( 93.98,290.71) --
	( 97.58,234.32) --
	(101.17,229.73) --
	(104.77,297.07) --
	(108.37,302.86) --
	(111.97,300.83) --
	(115.57,297.66) --
	(119.17,275.77) --
	(122.77,233.00) --
	(126.37,212.62) --
	(129.97,222.32) --
	(133.57,281.98) --
	(137.17,294.12) --
	(140.77,295.57) --
	(144.37,289.08) --
	(147.97,253.28) --
	(151.57,231.16) --
	(155.16,281.41) --
	(158.76,283.90) --
	(162.36,287.75) --
	(165.96,289.91) --
	(169.56,272.91) --
	(173.16,238.86) --
	(176.76,230.69) --
	(180.36,288.60) --
	(183.96,278.91) --
	(187.56,278.00) --
	(191.16,262.70) --
	(194.76,245.36) --
	(198.36,208.01) --
	(201.96,198.99) --
	(205.56,249.81) --
	(209.16,261.80) --
	(212.75,256.74) --
	(216.35,270.39) --
	(219.95,230.69) --
	(223.55,191.30) --
	(227.15,146.12) --
	(230.75,228.25) --
	(234.35,241.51) --
	(237.95,229.73) --
	(241.55,226.71) --
	(245.15,230.21) --
	(248.75,197.98) --
	(252.35,167.68) --
	(255.95,233.00) --
	(259.55,233.88) --
	(263.15,219.33) --
	(266.74,175.37) --
	(270.34,208.01) --
	(273.94,142.45) --
	(277.54,163.23) --
	(281.14,220.55) --
	(284.74,214.74) --
	(288.34,275.28) --
	(291.94,237.67) --
	(295.54,213.34) --
	(299.14,155.54) --
	(302.74,123.37) --
	(306.34,194.77) --
	(309.94,211.14) --
	(313.54,194.77) --
	(317.14,188.81) --
	(320.73,191.30) --
	(324.33,142.45) --
	(327.93,149.50) --
	(331.53,190.07) --
	(335.13,186.18) --
	(338.73,190.07) --
	(342.33,188.81) --
	(345.93,203.74) --
	(349.53,123.37) --
	(353.13, 70.49) --
	(356.73,235.61) --
	(360.33,224.02) --
	(363.93,208.81) --
	(367.53,186.18) --
	(371.13,169.74) --
	(374.73,158.26) --
	(378.32,138.42) --
	(381.92,177.10) --
	(385.52,196.93) --
	(389.12,178.75) --
	(392.72,158.26) --
	(396.32,146.12) --
	(399.92,123.37) --
	(403.52,123.37) --
	(407.12,202.83) --
	(410.72,178.75) --
	(414.32,167.68) --
	(417.92,163.23) --
	(421.52,129.01) --
	(425.12, 99.75) --
	(428.72,152.62) --
	(432.31,155.54) --
	(435.91,129.01) --
	(439.51,116.86) --
	(443.11,138.42) --
	(446.71, 87.61) --
	(450.31, 70.49) --
	(453.91,116.86) --
	(457.51, 99.75) --
	(461.11, 87.61) --
	(464.71,116.86);
	
	\path[draw=drawColor,line width= 0.4pt,line join=round,line cap=round] ( 65.18,257.25) circle (  2.25);
	
	\path[draw=drawColor,line width= 0.4pt,line join=round,line cap=round] ( 68.78,264.86) circle (  2.25);
	
	\path[draw=drawColor,line width= 0.4pt,line join=round,line cap=round] ( 72.38,206.35) circle (  2.25);
	
	\path[draw=drawColor,line width= 0.4pt,line join=round,line cap=round] ( 75.98,160.82) circle (  2.25);
	
	\path[draw=drawColor,line width= 0.4pt,line join=round,line cap=round] ( 79.58,278.15) circle (  2.25);
	
	\path[draw=drawColor,line width= 0.4pt,line join=round,line cap=round] ( 83.18,294.64) circle (  2.25);
	
	\path[draw=drawColor,line width= 0.4pt,line join=round,line cap=round] ( 86.78,299.37) circle (  2.25);
	
	\path[draw=drawColor,line width= 0.4pt,line join=round,line cap=round] ( 90.38,301.81) circle (  2.25);
	
	\path[draw=drawColor,line width= 0.4pt,line join=round,line cap=round] ( 93.98,290.71) circle (  2.25);
	
	\path[draw=drawColor,line width= 0.4pt,line join=round,line cap=round] ( 97.58,234.32) circle (  2.25);
	
	\path[draw=drawColor,line width= 0.4pt,line join=round,line cap=round] (101.17,229.73) circle (  2.25);
	
	\path[draw=drawColor,line width= 0.4pt,line join=round,line cap=round] (104.77,297.07) circle (  2.25);
	
	\path[draw=drawColor,line width= 0.4pt,line join=round,line cap=round] (108.37,302.86) circle (  2.25);
	
	\path[draw=drawColor,line width= 0.4pt,line join=round,line cap=round] (111.97,300.83) circle (  2.25);
	
	\path[draw=drawColor,line width= 0.4pt,line join=round,line cap=round] (115.57,297.66) circle (  2.25);
	
	\path[draw=drawColor,line width= 0.4pt,line join=round,line cap=round] (119.17,275.77) circle (  2.25);
	
	\path[draw=drawColor,line width= 0.4pt,line join=round,line cap=round] (122.77,233.00) circle (  2.25);
	
	\path[draw=drawColor,line width= 0.4pt,line join=round,line cap=round] (126.37,212.62) circle (  2.25);
	
	\path[draw=drawColor,line width= 0.4pt,line join=round,line cap=round] (129.97,222.32) circle (  2.25);
	
	\path[draw=drawColor,line width= 0.4pt,line join=round,line cap=round] (133.57,281.98) circle (  2.25);
	
	\path[draw=drawColor,line width= 0.4pt,line join=round,line cap=round] (137.17,294.12) circle (  2.25);
	
	\path[draw=drawColor,line width= 0.4pt,line join=round,line cap=round] (140.77,295.57) circle (  2.25);
	
	\path[draw=drawColor,line width= 0.4pt,line join=round,line cap=round] (144.37,289.08) circle (  2.25);
	
	\path[draw=drawColor,line width= 0.4pt,line join=round,line cap=round] (147.97,253.28) circle (  2.25);
	
	\path[draw=drawColor,line width= 0.4pt,line join=round,line cap=round] (151.57,231.16) circle (  2.25);
	
	\path[draw=drawColor,line width= 0.4pt,line join=round,line cap=round] (155.16,281.41) circle (  2.25);
	
	\path[draw=drawColor,line width= 0.4pt,line join=round,line cap=round] (158.76,283.90) circle (  2.25);
	
	\path[draw=drawColor,line width= 0.4pt,line join=round,line cap=round] (162.36,287.75) circle (  2.25);
	
	\path[draw=drawColor,line width= 0.4pt,line join=round,line cap=round] (165.96,289.91) circle (  2.25);
	
	\path[draw=drawColor,line width= 0.4pt,line join=round,line cap=round] (169.56,272.91) circle (  2.25);
	
	\path[draw=drawColor,line width= 0.4pt,line join=round,line cap=round] (173.16,238.86) circle (  2.25);
	
	\path[draw=drawColor,line width= 0.4pt,line join=round,line cap=round] (176.76,230.69) circle (  2.25);
	
	\path[draw=drawColor,line width= 0.4pt,line join=round,line cap=round] (180.36,288.60) circle (  2.25);
	
	\path[draw=drawColor,line width= 0.4pt,line join=round,line cap=round] (183.96,278.91) circle (  2.25);
	
	\path[draw=drawColor,line width= 0.4pt,line join=round,line cap=round] (187.56,278.00) circle (  2.25);
	
	\path[draw=drawColor,line width= 0.4pt,line join=round,line cap=round] (191.16,262.70) circle (  2.25);
	
	\path[draw=drawColor,line width= 0.4pt,line join=round,line cap=round] (194.76,245.36) circle (  2.25);
	
	\path[draw=drawColor,line width= 0.4pt,line join=round,line cap=round] (198.36,208.01) circle (  2.25);
	
	\path[draw=drawColor,line width= 0.4pt,line join=round,line cap=round] (201.96,198.99) circle (  2.25);
	
	\path[draw=drawColor,line width= 0.4pt,line join=round,line cap=round] (205.56,249.81) circle (  2.25);
	
	\path[draw=drawColor,line width= 0.4pt,line join=round,line cap=round] (209.16,261.80) circle (  2.25);
	
	\path[draw=drawColor,line width= 0.4pt,line join=round,line cap=round] (212.75,256.74) circle (  2.25);
	
	\path[draw=drawColor,line width= 0.4pt,line join=round,line cap=round] (216.35,270.39) circle (  2.25);
	
	\path[draw=drawColor,line width= 0.4pt,line join=round,line cap=round] (219.95,230.69) circle (  2.25);
	
	\path[draw=drawColor,line width= 0.4pt,line join=round,line cap=round] (223.55,191.30) circle (  2.25);
	
	\path[draw=drawColor,line width= 0.4pt,line join=round,line cap=round] (227.15,146.12) circle (  2.25);
	
	\path[draw=drawColor,line width= 0.4pt,line join=round,line cap=round] (230.75,228.25) circle (  2.25);
	
	\path[draw=drawColor,line width= 0.4pt,line join=round,line cap=round] (234.35,241.51) circle (  2.25);
	
	\path[draw=drawColor,line width= 0.4pt,line join=round,line cap=round] (237.95,229.73) circle (  2.25);
	
	\path[draw=drawColor,line width= 0.4pt,line join=round,line cap=round] (241.55,226.71) circle (  2.25);
	
	\path[draw=drawColor,line width= 0.4pt,line join=round,line cap=round] (245.15,230.21) circle (  2.25);
	
	\path[draw=drawColor,line width= 0.4pt,line join=round,line cap=round] (248.75,197.98) circle (  2.25);
	
	\path[draw=drawColor,line width= 0.4pt,line join=round,line cap=round] (252.35,167.68) circle (  2.25);
	
	\path[draw=drawColor,line width= 0.4pt,line join=round,line cap=round] (255.95,233.00) circle (  2.25);
	
	\path[draw=drawColor,line width= 0.4pt,line join=round,line cap=round] (259.55,233.88) circle (  2.25);
	
	\path[draw=drawColor,line width= 0.4pt,line join=round,line cap=round] (263.15,219.33) circle (  2.25);
	
	\path[draw=drawColor,line width= 0.4pt,line join=round,line cap=round] (266.74,175.37) circle (  2.25);
	
	\path[draw=drawColor,line width= 0.4pt,line join=round,line cap=round] (270.34,208.01) circle (  2.25);
	
	\path[draw=drawColor,line width= 0.4pt,line join=round,line cap=round] (273.94,142.45) circle (  2.25);
	
	\path[draw=drawColor,line width= 0.4pt,line join=round,line cap=round] (277.54,163.23) circle (  2.25);
	
	\path[draw=drawColor,line width= 0.4pt,line join=round,line cap=round] (281.14,220.55) circle (  2.25);
	
	\path[draw=drawColor,line width= 0.4pt,line join=round,line cap=round] (284.74,214.74) circle (  2.25);
	
	\path[draw=drawColor,line width= 0.4pt,line join=round,line cap=round] (288.34,275.28) circle (  2.25);
	
	\path[draw=drawColor,line width= 0.4pt,line join=round,line cap=round] (291.94,237.67) circle (  2.25);
	
	\path[draw=drawColor,line width= 0.4pt,line join=round,line cap=round] (295.54,213.34) circle (  2.25);
	
	\path[draw=drawColor,line width= 0.4pt,line join=round,line cap=round] (299.14,155.54) circle (  2.25);
	
	\path[draw=drawColor,line width= 0.4pt,line join=round,line cap=round] (302.74,123.37) circle (  2.25);
	
	\path[draw=drawColor,line width= 0.4pt,line join=round,line cap=round] (306.34,194.77) circle (  2.25);
	
	\path[draw=drawColor,line width= 0.4pt,line join=round,line cap=round] (309.94,211.14) circle (  2.25);
	
	\path[draw=drawColor,line width= 0.4pt,line join=round,line cap=round] (313.54,194.77) circle (  2.25);
	
	\path[draw=drawColor,line width= 0.4pt,line join=round,line cap=round] (317.14,188.81) circle (  2.25);
	
	\path[draw=drawColor,line width= 0.4pt,line join=round,line cap=round] (320.73,191.30) circle (  2.25);
	
	\path[draw=drawColor,line width= 0.4pt,line join=round,line cap=round] (324.33,142.45) circle (  2.25);
	
	\path[draw=drawColor,line width= 0.4pt,line join=round,line cap=round] (327.93,149.50) circle (  2.25);
	
	\path[draw=drawColor,line width= 0.4pt,line join=round,line cap=round] (331.53,190.07) circle (  2.25);
	
	\path[draw=drawColor,line width= 0.4pt,line join=round,line cap=round] (335.13,186.18) circle (  2.25);
	
	\path[draw=drawColor,line width= 0.4pt,line join=round,line cap=round] (338.73,190.07) circle (  2.25);
	
	\path[draw=drawColor,line width= 0.4pt,line join=round,line cap=round] (342.33,188.81) circle (  2.25);
	
	\path[draw=drawColor,line width= 0.4pt,line join=round,line cap=round] (345.93,203.74) circle (  2.25);
	
	\path[draw=drawColor,line width= 0.4pt,line join=round,line cap=round] (349.53,123.37) circle (  2.25);
	
	\path[draw=drawColor,line width= 0.4pt,line join=round,line cap=round] (353.13, 70.49) circle (  2.25);
	
	\path[draw=drawColor,line width= 0.4pt,line join=round,line cap=round] (356.73,235.61) circle (  2.25);
	
	\path[draw=drawColor,line width= 0.4pt,line join=round,line cap=round] (360.33,224.02) circle (  2.25);
	
	\path[draw=drawColor,line width= 0.4pt,line join=round,line cap=round] (363.93,208.81) circle (  2.25);
	
	\path[draw=drawColor,line width= 0.4pt,line join=round,line cap=round] (367.53,186.18) circle (  2.25);
	
	\path[draw=drawColor,line width= 0.4pt,line join=round,line cap=round] (371.13,169.74) circle (  2.25);
	
	\path[draw=drawColor,line width= 0.4pt,line join=round,line cap=round] (374.73,158.26) circle (  2.25);
	
	\path[draw=drawColor,line width= 0.4pt,line join=round,line cap=round] (378.32,138.42) circle (  2.25);
	
	\path[draw=drawColor,line width= 0.4pt,line join=round,line cap=round] (381.92,177.10) circle (  2.25);
	
	\path[draw=drawColor,line width= 0.4pt,line join=round,line cap=round] (385.52,196.93) circle (  2.25);
	
	\path[draw=drawColor,line width= 0.4pt,line join=round,line cap=round] (389.12,178.75) circle (  2.25);
	
	\path[draw=drawColor,line width= 0.4pt,line join=round,line cap=round] (392.72,158.26) circle (  2.25);
	
	\path[draw=drawColor,line width= 0.4pt,line join=round,line cap=round] (396.32,146.12) circle (  2.25);
	
	\path[draw=drawColor,line width= 0.4pt,line join=round,line cap=round] (399.92,123.37) circle (  2.25);
	
	\path[draw=drawColor,line width= 0.4pt,line join=round,line cap=round] (403.52,123.37) circle (  2.25);
	
	\path[draw=drawColor,line width= 0.4pt,line join=round,line cap=round] (407.12,202.83) circle (  2.25);
	
	\path[draw=drawColor,line width= 0.4pt,line join=round,line cap=round] (410.72,178.75) circle (  2.25);
	
	\path[draw=drawColor,line width= 0.4pt,line join=round,line cap=round] (414.32,167.68) circle (  2.25);
	
	\path[draw=drawColor,line width= 0.4pt,line join=round,line cap=round] (417.92,163.23) circle (  2.25);
	
	\path[draw=drawColor,line width= 0.4pt,line join=round,line cap=round] (421.52,129.01) circle (  2.25);
	
	\path[draw=drawColor,line width= 0.4pt,line join=round,line cap=round] (425.12, 99.75) circle (  2.25);
	
	\path[draw=drawColor,line width= 0.4pt,line join=round,line cap=round] (428.72,152.62) circle (  2.25);
	
	\path[draw=drawColor,line width= 0.4pt,line join=round,line cap=round] (432.31,155.54) circle (  2.25);
	
	\path[draw=drawColor,line width= 0.4pt,line join=round,line cap=round] (435.91,129.01) circle (  2.25);
	
	\path[draw=drawColor,line width= 0.4pt,line join=round,line cap=round] (439.51,116.86) circle (  2.25);
	
	\path[draw=drawColor,line width= 0.4pt,line join=round,line cap=round] (443.11,138.42) circle (  2.25);
	
	\path[draw=drawColor,line width= 0.4pt,line join=round,line cap=round] (446.71, 87.61) circle (  2.25);
	
	\path[draw=drawColor,line width= 0.4pt,line join=round,line cap=round] (450.31, 70.49) circle (  2.25);
	
	\path[draw=drawColor,line width= 0.4pt,line join=round,line cap=round] (453.91,116.86) circle (  2.25);
	
	\path[draw=drawColor,line width= 0.4pt,line join=round,line cap=round] (457.51, 99.75) circle (  2.25);
	
	\path[draw=drawColor,line width= 0.4pt,line join=round,line cap=round] (461.11, 87.61) circle (  2.25);
	
	\path[draw=drawColor,line width= 0.4pt,line join=round,line cap=round] (464.71,116.86) circle (  2.25);
	\end{scope}
	\begin{scope}
	\path[clip] (  0.00,  0.00) rectangle (505.89,361.35);
	\definecolor{drawColor}{RGB}{0,0,0}
	
	\path[draw=drawColor,line width= 0.4pt,line join=round,line cap=round] ( 61.58, 61.20) -- (421.52, 61.20);
	
	\path[draw=drawColor,line width= 0.4pt,line join=round,line cap=round] ( 61.58, 61.20) -- ( 61.58, 55.20);
	
	\path[draw=drawColor,line width= 0.4pt,line join=round,line cap=round] (133.57, 61.20) -- (133.57, 55.20);
	
	\path[draw=drawColor,line width= 0.4pt,line join=round,line cap=round] (205.56, 61.20) -- (205.56, 55.20);
	
	\path[draw=drawColor,line width= 0.4pt,line join=round,line cap=round] (277.54, 61.20) -- (277.54, 55.20);
	
	\path[draw=drawColor,line width= 0.4pt,line join=round,line cap=round] (349.53, 61.20) -- (349.53, 55.20);
	
	\path[draw=drawColor,line width= 0.4pt,line join=round,line cap=round] (421.52, 61.20) -- (421.52, 55.20);
	
	\node[text=drawColor,anchor=base,inner sep=0pt, outer sep=0pt, scale=  1.00] at ( 61.58, 39.60) {0};
	
	\node[text=drawColor,anchor=base,inner sep=0pt, outer sep=0pt, scale=  1.00] at (133.57, 39.60) {20};
	
	\node[text=drawColor,anchor=base,inner sep=0pt, outer sep=0pt, scale=  1.00] at (205.56, 39.60) {40};
	
	\node[text=drawColor,anchor=base,inner sep=0pt, outer sep=0pt, scale=  1.00] at (277.54, 39.60) {60};
	
	\node[text=drawColor,anchor=base,inner sep=0pt, outer sep=0pt, scale=  1.00] at (349.53, 39.60) {80};
	
	\node[text=drawColor,anchor=base,inner sep=0pt, outer sep=0pt, scale=  1.00] at (421.52, 39.60) {100};
	
	\path[draw=drawColor,line width= 0.4pt,line join=round,line cap=round] ( 49.20, 65.91) -- ( 49.20,276.95);
	
	\path[draw=drawColor,line width= 0.4pt,line join=round,line cap=round] ( 49.20, 65.91) -- ( 43.20, 65.91);
	
	\path[draw=drawColor,line width= 0.4pt,line join=round,line cap=round] ( 49.20,108.12) -- ( 43.20,108.12);
	
	\path[draw=drawColor,line width= 0.4pt,line join=round,line cap=round] ( 49.20,150.33) -- ( 43.20,150.33);
	
	\path[draw=drawColor,line width= 0.4pt,line join=round,line cap=round] ( 49.20,192.53) -- ( 43.20,192.53);
	
	\path[draw=drawColor,line width= 0.4pt,line join=round,line cap=round] ( 49.20,234.74) -- ( 43.20,234.74);
	
	\path[draw=drawColor,line width= 0.4pt,line join=round,line cap=round] ( 49.20,276.95) -- ( 43.20,276.95);
	
	\node[text=drawColor,rotate= 90.00,anchor=base,inner sep=0pt, outer sep=0pt, scale=  1.00] at ( 34.80, 65.91) {-1};
	
	\node[text=drawColor,rotate= 90.00,anchor=base,inner sep=0pt, outer sep=0pt, scale=  1.00] at ( 34.80,108.12) {0};
	
	\node[text=drawColor,rotate= 90.00,anchor=base,inner sep=0pt, outer sep=0pt, scale=  1.00] at ( 34.80,150.33) {1};
	
	\node[text=drawColor,rotate= 90.00,anchor=base,inner sep=0pt, outer sep=0pt, scale=  1.00] at ( 34.80,192.53) {2};
	
	\node[text=drawColor,rotate= 90.00,anchor=base,inner sep=0pt, outer sep=0pt, scale=  1.00] at ( 34.80,234.74) {3};
	
	\node[text=drawColor,rotate= 90.00,anchor=base,inner sep=0pt, outer sep=0pt, scale=  1.00] at ( 34.80,276.95) {4};
	
	\path[draw=drawColor,line width= 0.4pt,line join=round,line cap=round] ( 49.20, 61.20) --
	(480.69, 61.20) --
	(480.69,312.15) --
	( 49.20,312.15) --
	( 49.20, 61.20);
	\end{scope}
	\begin{scope}
	\path[clip] (  0.00,  0.00) rectangle (505.89,361.35);
	\definecolor{drawColor}{RGB}{0,0,0}
	
	\node[text=drawColor,anchor=base,inner sep=0pt, outer sep=0pt, scale=  1.20] at (264.94,332.61) {\bfseries Forecast for Awareness};
	
	\node[text=drawColor,anchor=base,inner sep=0pt, outer sep=0pt, scale=  1.00] at (264.94, 15.60) {Day of Summer};
	
	\node[text=drawColor,rotate= 90.00,anchor=base,inner sep=0pt, outer sep=0pt, scale=  1.00] at ( 10.80,186.67) {Normalized number of calls};
	\end{scope}
	\begin{scope}
	\path[clip] ( 49.20, 61.20) rectangle (480.69,312.15);
	\definecolor{drawColor}{RGB}{165,42,42}
	
	\path[draw=drawColor,line width= 0.4pt,line join=round,line cap=round] ( 65.18,257.25) --
	( 68.78,261.17) --
	( 72.38,241.09) --
	( 75.98,216.99) --
	( 79.58,233.28) --
	( 83.18,248.48) --
	( 86.78,260.54) --
	( 90.38,270.05) --
	( 93.98,274.74) --
	( 97.58,265.66) --
	(101.17,257.64) --
	(104.77,266.41) --
	(108.37,274.50) --
	(111.97,280.34) --
	(115.57,284.17) --
	(119.17,282.31) --
	(122.77,271.40) --
	(126.37,258.39) --
	(129.97,250.41) --
	(133.57,257.40) --
	(137.17,265.52) --
	(140.77,272.17) --
	(144.37,275.91) --
	(147.97,270.90) --
	(151.57,262.11) --
	(155.16,266.38) --
	(158.76,270.26) --
	(162.36,274.13) --
	(165.96,277.62) --
	(169.56,276.57) --
	(173.16,268.23) --
	(176.76,259.93) --
	(180.36,266.27) --
	(183.96,269.07) --
	(187.56,271.04) --
	(191.16,269.20) --
	(194.76,263.92) --
	(198.36,251.56) --
	(201.96,239.93) --
	(205.56,242.11) --
	(209.16,246.47) --
	(212.75,248.74) --
	(216.35,253.53) --
	(219.95,248.48) --
	(223.55,235.83) --
	(227.15,215.99) --
	(230.75,218.70) --
	(234.35,223.74) --
	(237.95,225.07) --
	(241.55,225.43) --
	(245.15,226.49) --
	(248.75,220.18) --
	(252.35,208.57) --
	(255.95,213.97) --
	(259.55,218.38) --
	(263.15,218.59) --
	(266.74,209.03) --
	(270.34,208.80) --
	(273.94,194.13) --
	(277.54,187.29) --
	(281.14,194.65) --
	(284.74,199.09) --
	(288.34,215.95) --
	(291.94,220.75) --
	(295.54,219.11) --
	(299.14,205.05) --
	(302.74,186.98) --
	(306.34,188.70) --
	(309.94,193.67) --
	(313.54,193.91) --
	(317.14,192.78) --
	(320.73,192.45) --
	(324.33,181.39) --
	(327.93,174.34) --
	(331.53,177.82) --
	(335.13,179.67) --
	(338.73,181.97) --
	(342.33,183.48) --
	(345.93,187.96) --
	(349.53,173.68) --
	(353.13,150.85) --
	(356.73,169.60) --
	(360.33,181.64) --
	(363.93,187.65) --
	(367.53,187.32) --
	(371.13,183.43) --
	(374.73,177.87) --
	(378.32,169.14) --
	(381.92,170.90) --
	(385.52,176.66) --
	(389.12,177.12) --
	(392.72,172.95) --
	(396.32,167.02) --
	(399.92,157.36) --
	(403.52,149.84) --
	(407.12,161.56) --
	(410.72,165.37) --
	(414.32,165.88) --
	(417.92,165.29) --
	(421.52,157.27) --
	(425.12,144.54) --
	(428.72,146.33) --
	(432.31,148.37) --
	(435.91,144.08) --
	(439.51,138.06) --
	(443.11,138.14) --
	(446.71,126.96) --
	(450.31,114.47) --
	(453.91,115.00) --
	(457.51,111.63) --
	(461.11,106.32) --
	(464.71,108.65);
	\definecolor{fillColor}{RGB}{165,42,42}
	
	\path[draw=drawColor,line width= 0.4pt,line join=round,line cap=round,fill=fillColor] ( 65.18,257.25) circle (  1.50);
	
	\path[draw=drawColor,line width= 0.4pt,line join=round,line cap=round,fill=fillColor] ( 68.78,261.17) circle (  1.50);
	
	\path[draw=drawColor,line width= 0.4pt,line join=round,line cap=round,fill=fillColor] ( 72.38,241.09) circle (  1.50);
	
	\path[draw=drawColor,line width= 0.4pt,line join=round,line cap=round,fill=fillColor] ( 75.98,216.99) circle (  1.50);
	
	\path[draw=drawColor,line width= 0.4pt,line join=round,line cap=round,fill=fillColor] ( 79.58,233.28) circle (  1.50);
	
	\path[draw=drawColor,line width= 0.4pt,line join=round,line cap=round,fill=fillColor] ( 83.18,248.48) circle (  1.50);
	
	\path[draw=drawColor,line width= 0.4pt,line join=round,line cap=round,fill=fillColor] ( 86.78,260.54) circle (  1.50);
	
	\path[draw=drawColor,line width= 0.4pt,line join=round,line cap=round,fill=fillColor] ( 90.38,270.05) circle (  1.50);
	
	\path[draw=drawColor,line width= 0.4pt,line join=round,line cap=round,fill=fillColor] ( 93.98,274.74) circle (  1.50);
	
	\path[draw=drawColor,line width= 0.4pt,line join=round,line cap=round,fill=fillColor] ( 97.58,265.66) circle (  1.50);
	
	\path[draw=drawColor,line width= 0.4pt,line join=round,line cap=round,fill=fillColor] (101.17,257.64) circle (  1.50);
	
	\path[draw=drawColor,line width= 0.4pt,line join=round,line cap=round,fill=fillColor] (104.77,266.41) circle (  1.50);
	
	\path[draw=drawColor,line width= 0.4pt,line join=round,line cap=round,fill=fillColor] (108.37,274.50) circle (  1.50);
	
	\path[draw=drawColor,line width= 0.4pt,line join=round,line cap=round,fill=fillColor] (111.97,280.34) circle (  1.50);
	
	\path[draw=drawColor,line width= 0.4pt,line join=round,line cap=round,fill=fillColor] (115.57,284.17) circle (  1.50);
	
	\path[draw=drawColor,line width= 0.4pt,line join=round,line cap=round,fill=fillColor] (119.17,282.31) circle (  1.50);
	
	\path[draw=drawColor,line width= 0.4pt,line join=round,line cap=round,fill=fillColor] (122.77,271.40) circle (  1.50);
	
	\path[draw=drawColor,line width= 0.4pt,line join=round,line cap=round,fill=fillColor] (126.37,258.39) circle (  1.50);
	
	\path[draw=drawColor,line width= 0.4pt,line join=round,line cap=round,fill=fillColor] (129.97,250.41) circle (  1.50);
	
	\path[draw=drawColor,line width= 0.4pt,line join=round,line cap=round,fill=fillColor] (133.57,257.40) circle (  1.50);
	
	\path[draw=drawColor,line width= 0.4pt,line join=round,line cap=round,fill=fillColor] (137.17,265.52) circle (  1.50);
	
	\path[draw=drawColor,line width= 0.4pt,line join=round,line cap=round,fill=fillColor] (140.77,272.17) circle (  1.50);
	
	\path[draw=drawColor,line width= 0.4pt,line join=round,line cap=round,fill=fillColor] (144.37,275.91) circle (  1.50);
	
	\path[draw=drawColor,line width= 0.4pt,line join=round,line cap=round,fill=fillColor] (147.97,270.90) circle (  1.50);
	
	\path[draw=drawColor,line width= 0.4pt,line join=round,line cap=round,fill=fillColor] (151.57,262.11) circle (  1.50);
	
	\path[draw=drawColor,line width= 0.4pt,line join=round,line cap=round,fill=fillColor] (155.16,266.38) circle (  1.50);
	
	\path[draw=drawColor,line width= 0.4pt,line join=round,line cap=round,fill=fillColor] (158.76,270.26) circle (  1.50);
	
	\path[draw=drawColor,line width= 0.4pt,line join=round,line cap=round,fill=fillColor] (162.36,274.13) circle (  1.50);
	
	\path[draw=drawColor,line width= 0.4pt,line join=round,line cap=round,fill=fillColor] (165.96,277.62) circle (  1.50);
	
	\path[draw=drawColor,line width= 0.4pt,line join=round,line cap=round,fill=fillColor] (169.56,276.57) circle (  1.50);
	
	\path[draw=drawColor,line width= 0.4pt,line join=round,line cap=round,fill=fillColor] (173.16,268.23) circle (  1.50);
	
	\path[draw=drawColor,line width= 0.4pt,line join=round,line cap=round,fill=fillColor] (176.76,259.93) circle (  1.50);
	
	\path[draw=drawColor,line width= 0.4pt,line join=round,line cap=round,fill=fillColor] (180.36,266.27) circle (  1.50);
	
	\path[draw=drawColor,line width= 0.4pt,line join=round,line cap=round,fill=fillColor] (183.96,269.07) circle (  1.50);
	
	\path[draw=drawColor,line width= 0.4pt,line join=round,line cap=round,fill=fillColor] (187.56,271.04) circle (  1.50);
	
	\path[draw=drawColor,line width= 0.4pt,line join=round,line cap=round,fill=fillColor] (191.16,269.20) circle (  1.50);
	
	\path[draw=drawColor,line width= 0.4pt,line join=round,line cap=round,fill=fillColor] (194.76,263.92) circle (  1.50);
	
	\path[draw=drawColor,line width= 0.4pt,line join=round,line cap=round,fill=fillColor] (198.36,251.56) circle (  1.50);
	
	\path[draw=drawColor,line width= 0.4pt,line join=round,line cap=round,fill=fillColor] (201.96,239.93) circle (  1.50);
	
	\path[draw=drawColor,line width= 0.4pt,line join=round,line cap=round,fill=fillColor] (205.56,242.11) circle (  1.50);
	
	\path[draw=drawColor,line width= 0.4pt,line join=round,line cap=round,fill=fillColor] (209.16,246.47) circle (  1.50);
	
	\path[draw=drawColor,line width= 0.4pt,line join=round,line cap=round,fill=fillColor] (212.75,248.74) circle (  1.50);
	
	\path[draw=drawColor,line width= 0.4pt,line join=round,line cap=round,fill=fillColor] (216.35,253.53) circle (  1.50);
	
	\path[draw=drawColor,line width= 0.4pt,line join=round,line cap=round,fill=fillColor] (219.95,248.48) circle (  1.50);
	
	\path[draw=drawColor,line width= 0.4pt,line join=round,line cap=round,fill=fillColor] (223.55,235.83) circle (  1.50);
	
	\path[draw=drawColor,line width= 0.4pt,line join=round,line cap=round,fill=fillColor] (227.15,215.99) circle (  1.50);
	
	\path[draw=drawColor,line width= 0.4pt,line join=round,line cap=round,fill=fillColor] (230.75,218.70) circle (  1.50);
	
	\path[draw=drawColor,line width= 0.4pt,line join=round,line cap=round,fill=fillColor] (234.35,223.74) circle (  1.50);
	
	\path[draw=drawColor,line width= 0.4pt,line join=round,line cap=round,fill=fillColor] (237.95,225.07) circle (  1.50);
	
	\path[draw=drawColor,line width= 0.4pt,line join=round,line cap=round,fill=fillColor] (241.55,225.43) circle (  1.50);
	
	\path[draw=drawColor,line width= 0.4pt,line join=round,line cap=round,fill=fillColor] (245.15,226.49) circle (  1.50);
	
	\path[draw=drawColor,line width= 0.4pt,line join=round,line cap=round,fill=fillColor] (248.75,220.18) circle (  1.50);
	
	\path[draw=drawColor,line width= 0.4pt,line join=round,line cap=round,fill=fillColor] (252.35,208.57) circle (  1.50);
	
	\path[draw=drawColor,line width= 0.4pt,line join=round,line cap=round,fill=fillColor] (255.95,213.97) circle (  1.50);
	
	\path[draw=drawColor,line width= 0.4pt,line join=round,line cap=round,fill=fillColor] (259.55,218.38) circle (  1.50);
	
	\path[draw=drawColor,line width= 0.4pt,line join=round,line cap=round,fill=fillColor] (263.15,218.59) circle (  1.50);
	
	\path[draw=drawColor,line width= 0.4pt,line join=round,line cap=round,fill=fillColor] (266.74,209.03) circle (  1.50);
	
	\path[draw=drawColor,line width= 0.4pt,line join=round,line cap=round,fill=fillColor] (270.34,208.80) circle (  1.50);
	
	\path[draw=drawColor,line width= 0.4pt,line join=round,line cap=round,fill=fillColor] (273.94,194.13) circle (  1.50);
	
	\path[draw=drawColor,line width= 0.4pt,line join=round,line cap=round,fill=fillColor] (277.54,187.29) circle (  1.50);
	
	\path[draw=drawColor,line width= 0.4pt,line join=round,line cap=round,fill=fillColor] (281.14,194.65) circle (  1.50);
	
	\path[draw=drawColor,line width= 0.4pt,line join=round,line cap=round,fill=fillColor] (284.74,199.09) circle (  1.50);
	
	\path[draw=drawColor,line width= 0.4pt,line join=round,line cap=round,fill=fillColor] (288.34,215.95) circle (  1.50);
	
	\path[draw=drawColor,line width= 0.4pt,line join=round,line cap=round,fill=fillColor] (291.94,220.75) circle (  1.50);
	
	\path[draw=drawColor,line width= 0.4pt,line join=round,line cap=round,fill=fillColor] (295.54,219.11) circle (  1.50);
	
	\path[draw=drawColor,line width= 0.4pt,line join=round,line cap=round,fill=fillColor] (299.14,205.05) circle (  1.50);
	
	\path[draw=drawColor,line width= 0.4pt,line join=round,line cap=round,fill=fillColor] (302.74,186.98) circle (  1.50);
	
	\path[draw=drawColor,line width= 0.4pt,line join=round,line cap=round,fill=fillColor] (306.34,188.70) circle (  1.50);
	
	\path[draw=drawColor,line width= 0.4pt,line join=round,line cap=round,fill=fillColor] (309.94,193.67) circle (  1.50);
	
	\path[draw=drawColor,line width= 0.4pt,line join=round,line cap=round,fill=fillColor] (313.54,193.91) circle (  1.50);
	
	\path[draw=drawColor,line width= 0.4pt,line join=round,line cap=round,fill=fillColor] (317.14,192.78) circle (  1.50);
	
	\path[draw=drawColor,line width= 0.4pt,line join=round,line cap=round,fill=fillColor] (320.73,192.45) circle (  1.50);
	
	\path[draw=drawColor,line width= 0.4pt,line join=round,line cap=round,fill=fillColor] (324.33,181.39) circle (  1.50);
	
	\path[draw=drawColor,line width= 0.4pt,line join=round,line cap=round,fill=fillColor] (327.93,174.34) circle (  1.50);
	
	\path[draw=drawColor,line width= 0.4pt,line join=round,line cap=round,fill=fillColor] (331.53,177.82) circle (  1.50);
	
	\path[draw=drawColor,line width= 0.4pt,line join=round,line cap=round,fill=fillColor] (335.13,179.67) circle (  1.50);
	
	\path[draw=drawColor,line width= 0.4pt,line join=round,line cap=round,fill=fillColor] (338.73,181.97) circle (  1.50);
	
	\path[draw=drawColor,line width= 0.4pt,line join=round,line cap=round,fill=fillColor] (342.33,183.48) circle (  1.50);
	
	\path[draw=drawColor,line width= 0.4pt,line join=round,line cap=round,fill=fillColor] (345.93,187.96) circle (  1.50);
	
	\path[draw=drawColor,line width= 0.4pt,line join=round,line cap=round,fill=fillColor] (349.53,173.68) circle (  1.50);
	
	\path[draw=drawColor,line width= 0.4pt,line join=round,line cap=round,fill=fillColor] (353.13,150.85) circle (  1.50);
	
	\path[draw=drawColor,line width= 0.4pt,line join=round,line cap=round,fill=fillColor] (356.73,169.60) circle (  1.50);
	
	\path[draw=drawColor,line width= 0.4pt,line join=round,line cap=round,fill=fillColor] (360.33,181.64) circle (  1.50);
	
	\path[draw=drawColor,line width= 0.4pt,line join=round,line cap=round,fill=fillColor] (363.93,187.65) circle (  1.50);
	
	\path[draw=drawColor,line width= 0.4pt,line join=round,line cap=round,fill=fillColor] (367.53,187.32) circle (  1.50);
	
	\path[draw=drawColor,line width= 0.4pt,line join=round,line cap=round,fill=fillColor] (371.13,183.43) circle (  1.50);
	
	\path[draw=drawColor,line width= 0.4pt,line join=round,line cap=round,fill=fillColor] (374.73,177.87) circle (  1.50);
	
	\path[draw=drawColor,line width= 0.4pt,line join=round,line cap=round,fill=fillColor] (378.32,169.14) circle (  1.50);
	
	\path[draw=drawColor,line width= 0.4pt,line join=round,line cap=round,fill=fillColor] (381.92,170.90) circle (  1.50);
	
	\path[draw=drawColor,line width= 0.4pt,line join=round,line cap=round,fill=fillColor] (385.52,176.66) circle (  1.50);
	
	\path[draw=drawColor,line width= 0.4pt,line join=round,line cap=round,fill=fillColor] (389.12,177.12) circle (  1.50);
	
	\path[draw=drawColor,line width= 0.4pt,line join=round,line cap=round,fill=fillColor] (392.72,172.95) circle (  1.50);
	
	\path[draw=drawColor,line width= 0.4pt,line join=round,line cap=round,fill=fillColor] (396.32,167.02) circle (  1.50);
	
	\path[draw=drawColor,line width= 0.4pt,line join=round,line cap=round,fill=fillColor] (399.92,157.36) circle (  1.50);
	
	\path[draw=drawColor,line width= 0.4pt,line join=round,line cap=round,fill=fillColor] (403.52,149.84) circle (  1.50);
	
	\path[draw=drawColor,line width= 0.4pt,line join=round,line cap=round,fill=fillColor] (407.12,161.56) circle (  1.50);
	
	\path[draw=drawColor,line width= 0.4pt,line join=round,line cap=round,fill=fillColor] (410.72,165.37) circle (  1.50);
	
	\path[draw=drawColor,line width= 0.4pt,line join=round,line cap=round,fill=fillColor] (414.32,165.88) circle (  1.50);
	
	\path[draw=drawColor,line width= 0.4pt,line join=round,line cap=round,fill=fillColor] (417.92,165.29) circle (  1.50);
	
	\path[draw=drawColor,line width= 0.4pt,line join=round,line cap=round,fill=fillColor] (421.52,157.27) circle (  1.50);
	
	\path[draw=drawColor,line width= 0.4pt,line join=round,line cap=round,fill=fillColor] (425.12,144.54) circle (  1.50);
	
	\path[draw=drawColor,line width= 0.4pt,line join=round,line cap=round,fill=fillColor] (428.72,146.33) circle (  1.50);
	
	\path[draw=drawColor,line width= 0.4pt,line join=round,line cap=round,fill=fillColor] (432.31,148.37) circle (  1.50);
	
	\path[draw=drawColor,line width= 0.4pt,line join=round,line cap=round,fill=fillColor] (435.91,144.08) circle (  1.50);
	
	\path[draw=drawColor,line width= 0.4pt,line join=round,line cap=round,fill=fillColor] (439.51,138.06) circle (  1.50);
	
	\path[draw=drawColor,line width= 0.4pt,line join=round,line cap=round,fill=fillColor] (443.11,138.14) circle (  1.50);
	
	\path[draw=drawColor,line width= 0.4pt,line join=round,line cap=round,fill=fillColor] (446.71,126.96) circle (  1.50);
	
	\path[draw=drawColor,line width= 0.4pt,line join=round,line cap=round,fill=fillColor] (450.31,114.47) circle (  1.50);
	
	\path[draw=drawColor,line width= 0.4pt,line join=round,line cap=round,fill=fillColor] (453.91,115.00) circle (  1.50);
	
	\path[draw=drawColor,line width= 0.4pt,line join=round,line cap=round,fill=fillColor] (457.51,111.63) circle (  1.50);
	
	\path[draw=drawColor,line width= 0.4pt,line join=round,line cap=round,fill=fillColor] (461.11,106.32) circle (  1.50);
	
	\path[draw=drawColor,line width= 0.4pt,line join=round,line cap=round,fill=fillColor] (464.71,108.65) circle (  1.50);
	\end{scope}
	\end{tikzpicture}
}
	\caption{The logarithmic plot of awareness (as measured by texts to ask for meal site locations) throughout the summer months. The open green dots are actual observations; the filled brown dots are the one step ahead forecast.}
	\label{fig:dlm}
\end{figure}
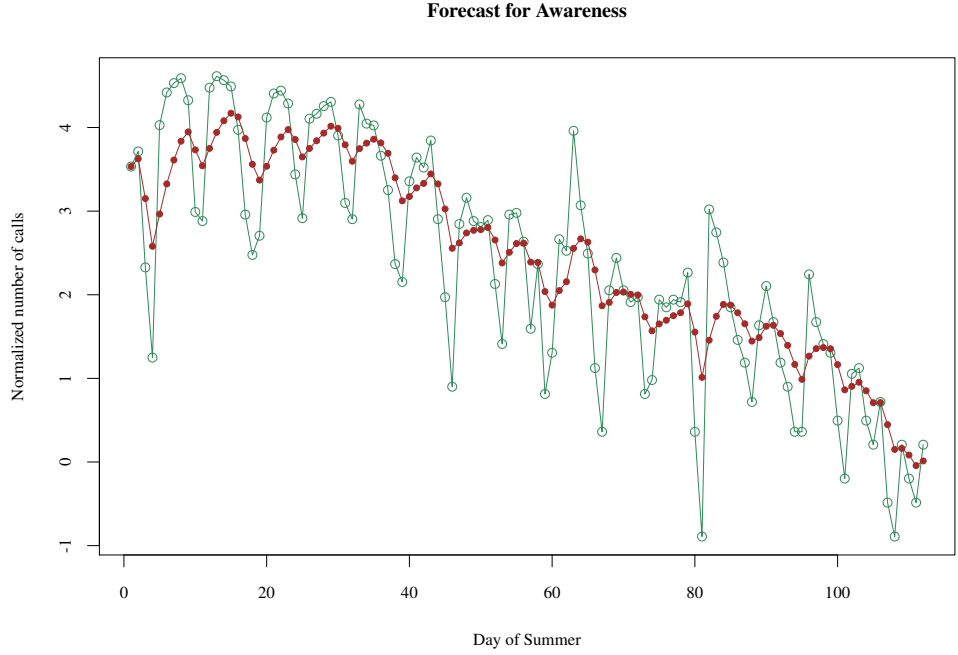

This one step ahead forecast factorizes according to the topology of the graph, allowing us to examine the plots of each of the terms. For our example, the one step ahead forecast factorizes: 

\begin{align*}
p(\bm{y}_t | \bm{y}^{t-1}) = \int_{\bm{\theta}_t(a)} p\{\bm{a}_t| \bm{a}^{t-1}\bm{\theta}_t(a)\} \, p\{\bm{\theta}_t(a)\} d\bm{\theta}_t(a) \\ 
\times \int_{\bm{\theta}_t(t)}p\{\bm{t}_t| \bm{a}^t, \bm{t}^{t-1},\bm{\theta}_t(t)\} \, p\{\bm{\theta}_t(t) | \bm{a}^{t-1},\bm{t}^{t-1}\} d\bm{\theta}_t(t) \\
\times \int_{\bm{\theta}_t(m)}p\{\bm{m}_t| \bm{\theta}_t(a),\bm{\theta}_t(t),\bm{m}^{t-1},\bm{\theta}_t(m)\} \, p\{\bm{\theta}_t(m) | \bm{a}^{t-1}, \bm{t}^{t-1}, \bm{m}^{t-1}\} d\bm{\theta}_t(m)
\end{align*}

Examining plots of the errors of each forecast can help determine what further structural adjustments should be made. For instance, in the Figure \ref{fig:dlm}, awareness has a cyclical nature, as people are less likely to text for an address of a meal site on weekends and holidays. This model can be adapted to include seasonal shifts using the equations from \cite{West1997}.

The implementation of this problem as an MDM rather than a BN maintains the strength of the relationships between each series and its regressors, respecting the natural language expression of the system by the domain experts. An additional feature of the MDM is that this representation renders the edges causal in the sense carefully argued in \cite{Queen2009}. For our model, note that while the two DAGs in Figure \ref{fig:mdm} both represent $A_t \condip M_t$, and are thus indistinguishable, the arrows in the MDM representation are unambiguous. The MDM offers a dynamic representation of a system in which the regressors influence a node contemporaneously. 

\subsection{Flow graph}
\label{sec:fg}
Structures can be adapted to meet additional constraints, such as conservation of a homogeneous mass transported in a system. However, these constraints motivate employing yet another graph with different semantics to transparently express the expert structural judgements. To illustrate how we might derive this from a natural language expression of a problem, consider the following example from the Summer Meals Program (SMP).

SMP provides no-cost meals to children under 18 at schools and community-based organisations during the summer months. SMP relies on food being procured from vendors, prepared by sponsors, and served at sites. Participation in the program is low, nationally 15\% percent of eligible children use the program \cite{Gundersen2011}. 
Sponsors, entities who provide and deliver meals, are reimbursed at a set rate per participant, but sponsors often struggle to break even. One of the key possible areas for cost cutting is the supply chain of the meals.  Community organisers hypothesize different interventions on each of these actors might help make the program more sustainable such as: 

\begin{itemize}
	\item A school district serving as a sponsor (Austin ISD) is having trouble breaking even. What happens when they partner with an external, more financially robust sponsor (City Square) to provide meals to the school. What is the effect on the supply chain of meals to the Elementary and Intermediate schools?
	\item Several smaller sponsors (among them the Boys and Girls Club) are having trouble breaking even and decide to create a collective to jointly purchase meals from a vendor (Revolution Foods). how does combining nodes in the network alter the flow of meals to the two Boys and Girls Club sites?
	\item Two sites, say apartment complexes A and B are low-performing, and the management decides to consolidate them. What is the long-term effect on a system?
	\item What happens when a sponsor, City Square, changes vendors from Revolution Foods to Aramark?
	\item What happens when one sponsor, Austin ISD, no longer administers the program and another sponsor, Boys and Girls Club takes responsibility for delivering food to the Intermediate and high Schools? 
\end{itemize}

Hearing the domain expert describe what types of intervention they would like to be able to model can elucidate the critical elements of the structure. In this example, the effect of the supply and transportation of meals through the network is key to the types of behaviour the modeller hopes to capture. This problem can be framed as a set quantity of meals moving through the system. Key model assumptions must always be checked with the domain expert. In this case, one of the key assumptions is that the number of children who are in need of meals and are likely to attend the program is relatively stable throughout the summer. This is a reasonable assumption, particularly when modelling a set population such as students in summer school or extracurricular programming. Community advocates verify that the assumption is reasonable because all of these sites and sponsors need a relatively set population in order to break even on the program. 

Additionally, to estimate the effect of the addition or removal of actors in the system, it is important to assume that the number of meals for children in need is conserved. Thus, if a sponsor and subsequent sites leave the program, then those children will access food at another sponsor's meal sites, provided transportation is available.  This assumption allows us to model particular interventions of interest, where combining, removing, or adding actors to the system is of particular interest. The dynamics of this particular problem involve the switching of ownership--what happens when the path flow of meals through the system changes--either a sponsor buys a meal from a different vendor, or a site turns to a different sponsor to supply their meals. This is a key component of the problem, but unfortunately it renders a key component of the problem intractable for the BN as shown below. However, Smith et al. discovered a methodology for re-framing this problem as a tractable variant of a BN that simultaneously remains faithful to the dynamics of the problem described above \cite{Smith2007}.

If we began modelling the process as a BN, we might begin by first identifying the actors involved.  A scenario for the key players in the city of Austin, Texas may consist of the following players at the vendor, sponsor, and site level. Levels are denoted by $z(i,j)$ where $i$ indicate the level (vendor, sponsor, or site), and $j$ differentiates between actors on a particular level.  In this example the players are:

\begin{itemize}
	\item $z(1,1)$ Revolution Foods
	\item $z(1,2)$ Aramark
	\item $z(2,1)$ City Square
	\item $z(2,2)$ Austin Independent School District 
	\item $z(2,3)$ Boys and Girls Club
	\item $z(3,1)$ Apartment complex A 
	\item $z(3,2)$ Apartment complex B 
	\item $z(3,3)$ Elementary School 
	\item $z(3,4)$ Intermediate School 
	\item $z(3,5)$ High School
	\item $z(3,6)$ Boys and Girls Club site A  
	\item $z(3,7)$ Boys and Girls Club site B
\end{itemize}

\begin{figure}[h]
	\centering
	\includegraphics[width=.7\textwidth]{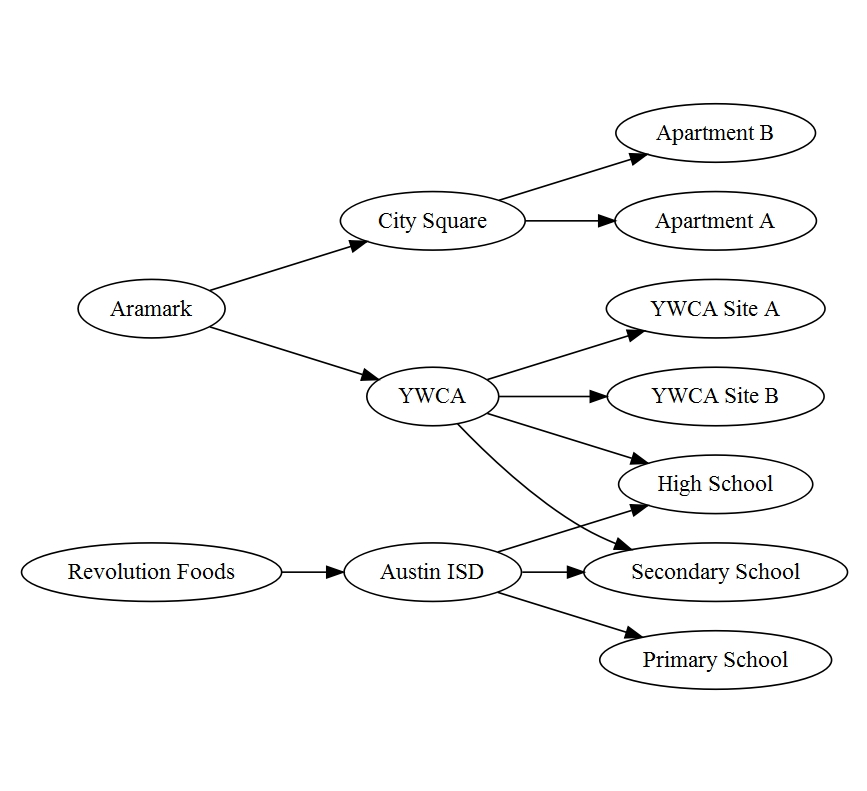}
	\caption{Flow Graph showing transfer of meals from vendors $z(1,j)$, to sponsors $z(2,j)$, to sites $z(3,j)$.}
	\label{fig:flow}
\end{figure}

These actors compose the nodes of the network; the edges represent the flow of meals between entities. For instance, vendor Aramark $z(1,2)$ prepares meals for sponsors at Austin ISD, $z(2,2)$, who in turn dispenses them at the Intermediate School, $z(3,4)$. We assume that each day, a set number of meals runs through the system. This list of actors can be readily obtained from natural language descriptions of the problem. Eliciting this information would simply requires the modeller to ask the domain expert to describe the flow of meals through each of the actors in the system. This structural elicitation and resultant graph in Figure \ref{fig:flow} are transparent to the expert, an advantage of customised modelling.

As the modeller begins to check the relationships encoded in the graphical model elicited in Figure \ref{fig:flow}, the missing edges between actors in a given level means that each of the sponsors is unaffected by the meals being transported to and from the other sponsors. However, this is not realistic for closed sites because the experts have told us that knowing the number of meals served at all but one sponsor gives us perfect information about the remaining sponsor, as we know the number of meals served by sponsors remains constant! For instance, if we know how many meals are prepared by Aramark, $z(1,1)$, then we have perfect information about how many are prepared by Revolution Foods, $z(1,2)$, because meals are conserved at each level, implying a directed line from $z(1,1)$ to $z(1,2)$. Modelling this process graphically, as in Figure \ref{fig:flow}, induces severe dependencies in the network if we consider the graph to be a BN. Thus, the problem as the experts have expressed it cannot be represented as a BN. 

However, by decomposing the information in Figure \ref{fig:flow} to into paths as shown in \cite{Smith2007}, we can apply the methodology of dynamic Bayesian networks. Denote $\bm{\phi}_t'[l] = (\phi_t(l,1), \phi_t(l,2), \ldots, \phi_t(l,n_l)  )$ as the node states vector for each of the three levels, where $\phi_t(l,j_l)$ represents the mass owned by player $z(l,j_l)$ during time $t$. This probabilistic representation allows the modeller to retain the advantages of the clear representation in Figure \ref{fig:flow} to draw information about the system from the experts as well as the computational convenience of the BN machinery. 

The full methodology for translating the hierarchical flow graph to the DBN representation is given in \cite{Smith2007}, here we simply state the elements of the model that would need to be a part of the probability elicitation. Information about the numbers of meals held by each entity at each day during the summer can be represented by a time series vector $\bm{X}_t' = (\bm{X}_t'[1], \bm{X}_t'[2], \bm{X}_t'[3] )$, representing the number of meals at the vendor, sponsor, and site levels respectively. Next, we represent the paths of meals travelling from vendor to meal site as aggregates of the product amounts. The paths in this diagram are: 
\begin{align}
\pi(1) &= \{ z(1,1), z(2,1), z(3,2) \} \qquad \pi(2) = \{ z(1,1), z(2,1), z(3,1) \} \\ \nonumber
\pi(3) &= \{ z(1,1), z(2,3), z(3,6) \} \qquad \pi(4) = \{ z(1,1), z(2,3), z(3,7) \} \\ \nonumber
\pi(5) &= \{ z(1,1), z(2,3), z(3,5) \} \qquad \pi(6) =\{ z(1,1), z(2,3), z(3,4) \} \\ \nonumber
\pi(7) &= \{ z(1,2), z(2,2), z(3,5) \} \qquad \pi(8) =\{ z(1,2), z(2,2), z(3,4) \} \\ \nonumber
\pi(9) &= \{ z(1,2), z(2,2), z(3,3) \} 
\end{align}

Fully embellishing this model involves eliciting the core states, the underlying drivers of the number of meals passing through each of the actors. These can be readily adapted to reflect the beliefs of different domain experts. For instance, different school districts often follow different summer school schedules, so if the advocates were interested in applying the model to a different region, it would simply require updating the core state parameters. The information about the path flows is most readily supplied through available data about the number of meals prepared, transported, and served throughout the summer.

\section{Discussion}
\label{sec:disc}
The case studies in Section \ref{sec:foodex} show how drawing the structure from the experts' natural langauge description motivates the development of more flexible models that can highligh key features of a domain problem. The SBP example shows that a BN is appropriate when the expert describes a problem as a set of elements that depend on each other. The SNAP application example highlights the advantages of a tree-based approach when the experts describe a series of events and outcomes. The open SMP example shows how additional restrictions on the BN structure can draw out the contemporaneous strengths between elements of the model that is crucial to the experts' description. Lastly, the flow of meals in a system shows how working with the accessible representation of meal flow in a system can be trnaslated into a valid structure while remaining faithful to the assumptions expressed by the expert. 

A summary table is shown in Table \ref{tab:exs} citing additional examples of applications of these bespoke graphical models we have used in the past. References are given for two classes of models, chain graphs and regulatory graphs that are not explored in this paper. This is of course a small subset of all the formal graphical frameworks now available. These case studies and applications in the table are examples from the toolkit of customising models. 

\begin{table}
		\label{tab:exs}
	
	\begin{tabular}{p{2cm}p{3cm}p{3cm}p{3cm}}
		\hline\noalign{\smallskip}
		Name & Description & When to use & Applications \\
		\noalign{\smallskip}\svhline\noalign{\smallskip}
(Dynamic) Bayesian Network&Directed acyclic graph of random variables& Systems naturally expressed as dependence structure between random variables & Biological networks \cite{JQSdecisionAnalysis}, ecological conservation \cite{Korb2009}\\ \hline
(Dynamic) Chain Event Graph&Derived from event tree coloured to represent conditional independence& Asymmetric problems, problem description is told as a series of unfolding events&Healthcare outcomes \cite{Barclay2014}, forensic evidence \cite{CEGbook} \\\hline
Chain Graphs &Hybrid graph with directed and undirected edges&Problem description has both directional and ambiguous relationships& Mental health \cite{Cox1993}, social processes \cite{Cox2000} \\\hline
Flow Graph&Hierarchical flow network& Supply and demand problems, homogeneous flows &Commodity supply \cite{Smith2007}\\\hline
Multi-regression Dynamic model& Collection of regressions where the parents are the regressors & Contemporaneous effects between time series&Marketing\cite{Smith1993}, traffic flows\cite{Queen2009}, neural fMRI activity\cite{Costa2015} \\\hline
Regulatory Graph & Graph customised to regulatory hypotheses& Need to test a regulatory hypothesis&Biological control mechanisms \cite{Liverani2015}\\
		\noalign{\smallskip}\hline\noalign{\smallskip}
	\end{tabular}
\end{table}

Generally, allowing these representations to capture dynamics unique to a given application cultivates more suitable representations. Just as the $d$-separation theorem allows us to reason about conditional independence in the BN, analogous theorems elucidate the dependence structure of custom representations. Each of these examples of elicited structure has its own logic which can be verified by examining the conditional independence statements and confirming with the expert that the model accurately conveys the expert's beliefs. 

\begin{figure}[h]	
	\centering
	\includegraphics[width=.95\textwidth]{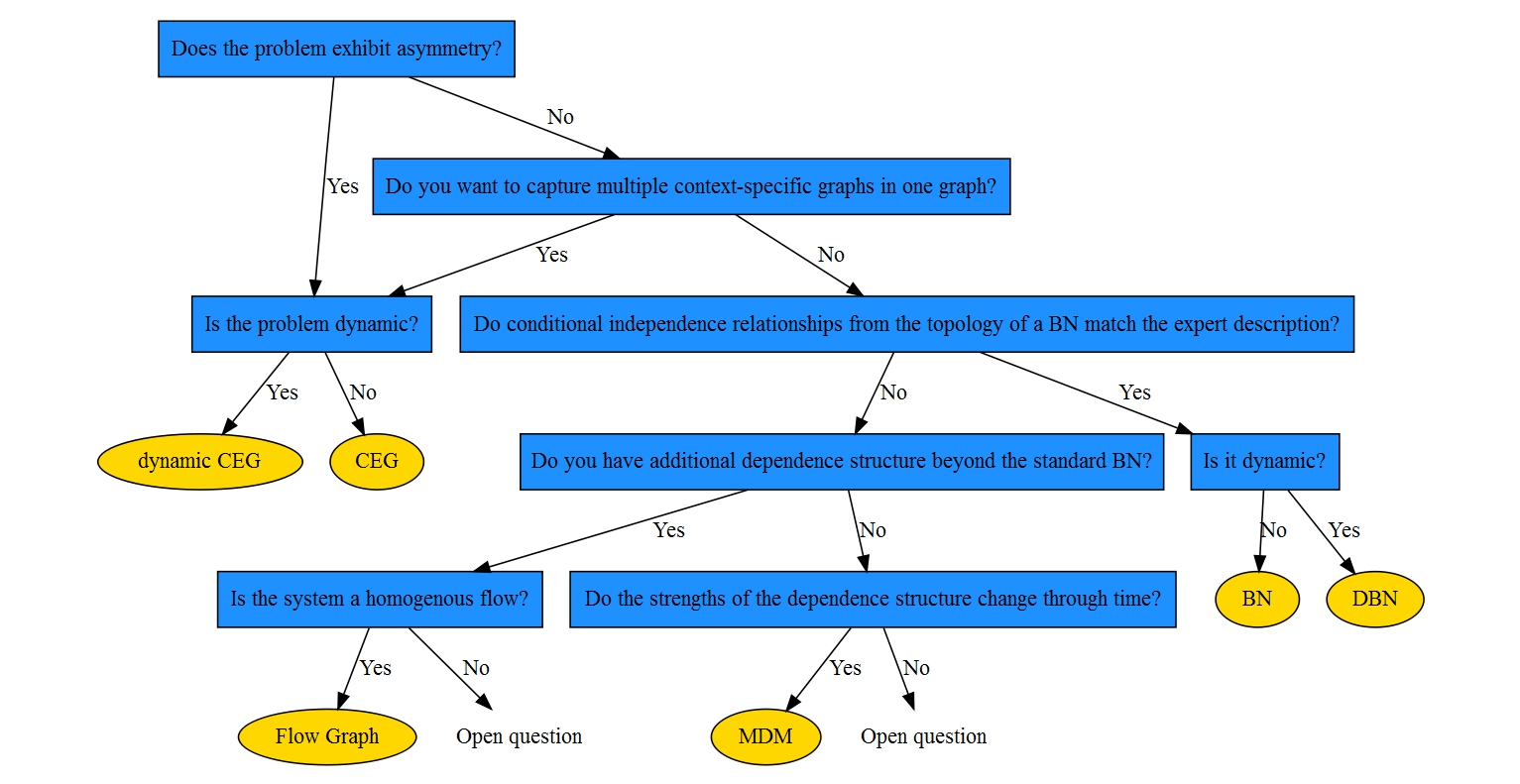}
	\caption{Flow chart to guide picking an appropriate structure of the ones discussed.}
	\label{fig:pick}
\end{figure}

Carefully drawing structure from an expert's natural language description is not an exact science. We have offered a few guidelines for when to use particular models summarised in the flow chart in Figure \ref{fig:pick}. The examples discussed here are far from exhaustive and Figure \ref{fig:pick} also highlights areas of open research. Spirtes et al. \cite{Spirtes2016} confirms that determining what new classes of models might be more appropriate than a BN for a given domain. A full protocol for choosing one customising model over another remains to be formalised. While software for BN elicitation is ubiquitous, robust software for these alternative models is under development.

The premise of drawing the structure from a natural language description rather than tweaking a model to fit an existing structure represents a substantial shift in how modellers elicit structure. Furthermore, inference on each of these novel engenders customised notions of causation, as the full probability representations of customised models each admits its own causal algebras. The causal effects following intervention in a BN are well studied, and these methods can be extended to custom classes of models discussed here. A thorough investigation of causal algebras is beyond the scope of this paper, but it offer further motivation for careful attention to structure in the elicitation process. In a later paper, we will demonstrate how each structural class has its own causal algebra and that for causation to be meaningful the underlying structure on which it is based needs to properly reflect domain knowledge. The work of customised structure elicitation is a relatively poorly explored space. We hope this paper excites others to develop new tools to make problem descriptions more powerful and reliable.

\bibliographystyle{plain}

\bibliography{Structure}
\end{document}